\documentclass[acp, manuscript]{copernicus}

\renewcommand{\thesubsection}{\thesection.\alph{subsection}}

\usepackage{latexsym,euscript,textcomp}
\usepackage{color,graphics,epsf,dcolumn}
\usepackage{epstopdf,ulem}
\usepackage{blindtext}
\usepackage{threeparttable}

\begin{document}
\nolinenumbers

\title{Alternative expression for the maximum potential intensity of tropical cyclones}


\Author[1,2]{Anastassia M.}{Makarieva}
\Author[1]{Andrei V.}{Nefiodov}

\affil[1]{Theoretical Physics Division, Petersburg Nuclear Physics Institute, Gatchina  188300, St.~Petersburg, Russia}
\affil[2]{Institute for Advanced Study, Technical University of Munich,  Garching 85748, Germany}


\runningtitle{Alternative expression for maximum intensity}

\runningauthor{Makarieva et al.}

\correspondence{A. M. Makarieva (ammakarieva@gmail.com)}

\received{}
\pubdiscuss{} 
\revised{}
\accepted{}
\published{}


\firstpage{1}

\maketitle

\begin{abstract}
Emanuel{\textquoteright}s concept of maximum potential intensity (E-PI) estimates the maximum velocity of tropical cyclones from environmental parameters. At the point of maximum wind, E-PI{\textquoteright}s key equation relates proportionally the centrifugal acceleration (squared maximum velocity divided by radius) to the radial gradient of saturated moist entropy. The proportionality coefficient depends on the outflow temperature. Here it is shown that a different relationship between the same quantities derives straightforwardly from the gradient-wind balance and the definition of entropy, with the proportionality coefficient depending on the radial gradient of local air temperature. The robust alternative  reveals a previously unexplored constraint: for E-PI to be valid, the outflow temperature should be a function of the radial temperature gradient at the point of maximum wind. When the air is horizontally isothermal (which, as we argue,  is not an uncommon condition), this constraint cannot be satisfied, and E-PI{\textquoteright}s key equation underestimates the squared maximum velocity by approximately twofold. This explains {\textquotedblleft}superintensity{\textquotedblright} (maximum wind speeds exceeding E-PI). The new formulation predicts less superintensity at higher tem\-pe\-ra\-tu\-res, corroborating recent numerical simulations. Previous analyses are re-evaluated to reveal inconsistent support for the explanation of superintensity by supergradient winds alone. In Hurricane Isabel 2003, maximum superintensity is found to be associated with minimal gradient-wind imbalance. Modified to diagnostically account for supergradient winds, the new formulation shows that air temperature increasing towards the storm center can mask the effect of gradient-wind imbalance, thus reducing {\textquotedblleft}superintensity{\textquotedblright} and formally bringing E-PI closer to observations. The implications of these findings for assessing real storms are discussed.
\end{abstract}

\introduction  
\label{intr}

Tropical storms threaten human lives and livelihoods. Numerical models can simulate a wide range of storm intensities under the same environmental conditions \citep[e.g.,][]{tao20}. Thus it is desirable to  have a reliable theoretical framework that would, from the first principles, confine model outputs to the domain of reality \citep{emanuel20}.  The theoretical formulation for maximum potential intensity of tropical cyclones by \citet{em86} (E-PI) has been long considered as an approximate upper limit on storm intensity (see discussions by \citet{garner15}, \citet{kieu2016} and \citet[][]{kowaleski16}). At the same time, the phenomenon of  {\textquotedblleft}superintensity{\textquotedblright}, when the observed or modelled storm velocities exceed E-PI, has been perceived as an important research challenge \citep[e.g.,][]{persing2003,montgomery06,bryan09b,re19,Li20c}. Since the strongest storms are the most dangerous ones, it is important to understand when and why the theoretical limits can be exceeded. The principal way of approaching the superintensity problem was to reveal how the E-PI assumptions can be modified to yield greater intensities.  For example, \citet{montgomery06} suggested that superintensity could result from an additional heat source provided by the storm eye (a source of energy not considered in E-PI). \citet{bryan09c} evaluated this mechanism in a numerical modelling study and found it to be small. In another numerical simulation, \citet{bryan09b} investigated how superintensity could result from the flow being supergradient (while E-PI assumed the gradient-wind balance) and found this mechanism to be more significant than the eye heat source. For a recent overview of superintensity assessments in modelling studies see \citet{re19}.

Here we present a different approach. We show that, even in the case when all the E-PI assumptions hold, E-PI will systematically underestimate storm intensities provided the air is horizontally isothermal at the point of maximum wind.  This conclusion follows straightforwardly from the definition of saturated moist entropy.  At the point of maximum wind E-PI relates the radial gradients $\partial s^*/\partial r$ and $\partial p/\partial r$ of saturated moist entropy $s^*$ and air pressure $p$ via an external parameter (the outflow temperature $T_o$). However, $s^*$ being a state variable, its radial gradient is a local function of the radial gradients of air pressure $p$ and temperature $T$. Thus, specifying a relationship between $\partial s^*/\partial r$ and $\partial p/\partial r$ uniquely sets $\partial T/\partial r$. Conversely, setting $\partial T/\partial r = 0$  relates $\partial s^*/\partial r$ and $\partial p/\partial r$ in a specific way that, under common atmospheric conditions, is shown here to be incompatible with E-PI. For E-PI to match observations, $\partial T/\partial r$ at the point of maximum wind must be a function of the outflow temperature $T_o$.

The assumption of horizontal isothermy is not required for the derivation of E-PI{\textquoteright}s expression for maximum velocity (see, e.g., Eqs.~(1)-(22) of \citet{emanuel11} and Eq.~(\ref{vE}) below). Probably because of that, and since the above constraint on E-PI remained unknown, it became common, in diverse E-PI developments, to treat the horizontal temperature gradient as a free parameter to be specified at one{\textquoteright}s discretion, with a common assumption of $\partial T/\partial r = 0$ at the top of the boundary layer. In particular, \citet[][p.~2245]{emanuel11} assumed temperature $T_b$ of the top of the boundary layer to be constant in their attempt to modify E-PI{\textquoteright}s expression for maximum  velocity to account for a possible dependence of the outflow temperature on angular momentum. \citet[][p.~588]{em86}, see also \citet{em88}, likewise assumed a constant $T_b$ while assessing storm intensity in terms of the minimum central pressure. The same assumption was used in the many E-PI representations of tropical cyclones as Carnot heat engines (e.g., \citet[][their Fig.~13]{em86}, \citet[][their Eq.~(7)]{em91}, \citet{emanuel2006}).  More recently, \citet{re19} extended the E-PI concept to describe surface winds and also assumed horizontal isothermy. Thus, $\partial T/\partial r = 0$ appears to be an {\it a priori} plausible and widely used approximation, which we will therefore consider in greater detail.

To clarify the logic of the foregoing analyses for our readers, we believe that it could be useful to contrast the following two viewpoints. The first one is from one of our reviewers: E-PI does not make any assumptions about the horizontal temperature gradient; E-PI generally conforms to observations and numerical simulations, and where it does not, the mismatch can be largely explained by supergradient winds. As far as E-PI does not assume horizontal isothermy, it is not valid to consider the case of horizontal isothermy and to infer from this, 
as we do based on our alternative formulation, that E-PI underestimates maximum winds.

Our alternative viewpoint is as follows.  While we agree that E-PI does not {\it assume} anything about horizontal temperature gradient,  here we show that it {\it predicts} it. Thus, assessing the value of $\partial T/\partial r$  is a test of E-PI{\textquoteright}s validity.  We will give examples and argue that in many cases a negligible or small temperature gradient at the point of maximum wind is a plausible approximation. In those cases E-PI does not conform to observations and does underestimate maximum winds. We have re-evaluated the study of supergradient winds by \citet{bryan09b} to reveal that their analysis is not self-consistent and does not explain superintensity either in Hurricane Isabel 2003 or in their own numerical model (see, respectively, section~\ref{isab} and appendix~\ref{bryan}). At this point, this leaves the new formulation, easily modifiable to diagnostically account for supergradient winds, the only available explanation of {\textquotedblleft}superintensity{\textquotedblright}.

We derive the alternative expression for maximum potential intensity and discuss how the conventional and alternative expressions relate to each other as dependent on temperature in section~\ref{epi}. We illustrate the obtained relationships with the data for Hurricane Isabel 2003 in section~\ref{isab} and discuss their implications in section~\ref{concl}.

\section{Different expressions for maximum intensity}
\label{epi}
\subsection{Conventional E-PI}

E-PI has three blocks, with distinct sets of assumptions applied to each block: the free troposphere including the top of the boundary layer, the interior of the boundary layer and the ocean-atmosphere interface (Table~\ref{tabEPI}). Here we focus on the first block.

For the free troposphere, the key relationship of E-PI is between saturated moist entropy $s^*$  and angular momentum $M$ \citep[for a compact derivation see][their Eq.~(11)]{emanuel11}:
\noindent
\begin{equation}
\label{dsE}
-(T_1 - T_2)\frac{ds^*}{dM} = \frac{v_1}{r_1} - \frac{v_2}{r_2}, \quad z \ge z_b.
\end{equation}
Here $T_1$, $T_2$  and $v_1$, $v_2$ are, respectively, air temperatures and tangential wind speeds at arbitrary distances $r_1$ and $r_2$ from the storm center on a surface defined by the given value of $ds^*/dM$, $z_b$ is the height of the boundary layer, and
\noindent
\begin{equation}\label{M}
M = vr + \frac{1}{2}fr^2.
\end{equation}
The Coriolis parameter $f \equiv 2 \Omega \sin \varphi$ is assumed constant ($\varphi$ is latitude, $\Omega$ is the angular velocity of Earth{\textquoteright}s rotation). For the definition of saturated moist entropy see Eq.~(\ref{s}) in appendix~\ref{altf}.

Relationship (\ref{dsE}) is derived assuming that for $z\ge z_b$ hydrostatic and gradient-wind balances hold and surfaces of constant $s^*$ and $M$ coincide \citep{em86,emanuel11}.

One can choose $r_1$ at the top of the boundary layer ($z = z_b$) and $r_2$ in the outflow in the free troposphere, where 
$v_2 = 0$. Then, since $s^* = s^*(M)$, multiplying Eq.~(\ref{dsE}) by $\partial M/\partial r$ gives (cf. Eq.~(12) of  \citet{em86}):
\begin{equation}\label{dsE2}
\varepsilon T_b \frac{\partial s^*}{\partial r} =   -\frac{v}{r}\frac{\partial M}{\partial r}, \quad z = z_b.
\end{equation}
Here $\varepsilon \equiv (T_b - T_o)/T_b$ is the Carnot efficiency, $T_b=T_1$ is the local temperature at the top of the boundary layer, $T_o=T_2$ is the outflow temperature. Note that Eq.~(\ref{dsE2}) does not assume $\partial T_b/\partial r = \rm const$.

If the radial gradients of $s^*$ and $M$ relate as their respective surface fluxes $\tau_s$ and $\tau_M$, see Table~\ref{tabEPI}, 
\begin{equation}\label{tau}
\frac{\partial s^*/\partial r}{\partial M/\partial r} = \frac{\tau_s}{\tau_M} = -\frac{C_k}{C_D}\frac{k_s^* - k_0}{T_s r v_s},
\end{equation}
Eq.~(\ref{dsE2}) yields the E-PI expression for maximum intensity
\begin{equation}\label{vE}
v_E^2 = \varepsilon \frac{C_k}{C_D}(k_s^* - k_0).
\end{equation}
All the notations are given in Table~\ref{tabEPI}. The local difference between saturated $k_s^*$ and actual $k_0$ enthalpies at the air-sea interface is {\it a priori} unknown. To relate it to environmental parameters, yet another set of assumptions is required, see block E-III in Table~\ref{tabEPI}.

Equation~(\ref{vE}) assumes that $v(z_b) = v_s$ and $T_b = T_s$, where the subscript $s$ refers to $z=0$. Nuances stemming from $v(z_b) \ne v_s$ and $T_b \ne T_s$ \citep[for their discussion, see][]{emanuel11} do not matter for our foregoing results, as these nuances equally affect $v_E$ and the alternative estimate $v_A$ to be derived in the next section.

\subsection{Alternative expression for maximum potential intensity}

Since saturated moist entropy $s^*$ is a state variable, its radial gradient can be expressed in terms of the radial gradients of air pressure and temperature (see Eq.~\eqref{Tds3}):
\begin{equation}\label{sr}
\frac{1}{1+\zeta}T \frac{\partial s^*}{\partial r} = -\alpha_d\frac{\partial p}{\partial r}\left(1 - \frac{1}{\Gamma}\frac{\partial T/\partial r}{\partial p/\partial r}\right),
\end{equation}
where $p$ is air pressure, $\zeta \equiv L \gamma_d^*/(R T)$, $R = 8.3$~J~mol$^{-1}$~K$^{-1}$ is the universal gas constant, $\gamma_d^* \equiv p_v^*/p_d$, $p_v^*$ is the  partial pressure of saturated water vapor, $p_d$ is the partial pressure of dry air, $L \simeq 45$~kJ~mol$^{-1}$ is the latent heat of vaporization, $\Gamma$ (K Pa$^{-1}$) is the moist adiabatic lapse rate of air temperature (see its definition \eqref{Gm}), and $\alpha_d \equiv 1/\rho_d$ is the inverse dry air density. Below we assume $\alpha_d \simeq \alpha$, where $\alpha \equiv 1/\rho$ is the inverse air density. Equation (\ref{sr}) does not contain any assumptions but follows directly from the definition of saturated moist entropy. 

In gradient-wind balance, 
\noindent
\begin{equation}\label{gwb}
\alpha \frac{\partial p}{\partial r} = \frac{v^2}{r} + fv,
\end{equation}
and at the radius of maximum wind $r = r_m$, where $\partial v/\partial r = 0$ and $\partial M/\partial r = v + fr$,
we have
\noindent
\begin{equation}\label{dpr}
\alpha \frac{\partial p}{\partial r} = \frac{v}{r} \frac{\partial M}{\partial r}, \,\,\,r = r_m.
\end{equation}
Introducing
\begin{equation}\label{C}
\mathcal{C} \equiv 1 -\frac{1}{\Gamma} \frac{\partial T/\partial r}{\partial p/\partial r}
\end{equation}
and combining Eqs.~(\ref{sr})-(\ref{C}) 
we obtain an alternative version of Eq.~(\ref{dsE2}),
\begin{equation} 
\label{alt1}
\frac{1}{\mathcal{C}(1+\zeta)} T_b \frac{\partial s^*}{\partial r} =   -\frac{v}{r}\frac{\partial M}{\partial r}, \quad  r = r_m,  z = z_b, 
\end{equation}
from which, using Eq.~(\ref{tau}), our alternative expression for maximum intensity results:
\begin{equation}\label{vA}
v_A^2 = \frac{1}{\mathcal{C} (1+ \zeta)}\frac{C_k}{C_D}(k_s^* - k_0).
\end{equation}

Equation~(\ref{tau}) has been used for deriving maximum intensities (\ref{vE}) and (\ref{vA}) from, respectively, Eqs.~(\ref{dsE2}) and (\ref{alt1}).  The assumptions yielding Eq.~(\ref{tau}) pertain to the boundary layer. They are independent of the E-PI assumptions behind Eq.~(\ref{dsE2}) that pertain to the free troposphere, see Table~\ref{tabEPI}. The difference in maximum intensities $v_E$ (\ref{vE}) and $v_A$ (\ref{vA}),
\begin{equation}\label{diff}
\left(\frac{v_A}{v_E}\right)^2 = \frac{1}{\varepsilon \mathcal{C} (1+ \zeta)},
\end{equation}
stems from Eqs.~(\ref{dsE2}) and (\ref{alt1}). Both equations assume gradient-wind balance. Equation~(\ref{dsE2}) is valid at $z = z_b$, Eq.~(\ref{alt1}) is valid at the point of maximum wind $r = r_m$, $z = z_b$. (We assume, as does E-PI, that the point of maximum wind is at $z = z_b$.)  Equation~(\ref{dsE2}) assumes hydrostatic balance and $s^* = s^*(M)$ for $z \ge z_b$. Equation~(\ref{alt1}) does not require these assumptions. Therefore, at the point of maximum wind, Eq.~(\ref{alt1}) is more general than Eq.~(\ref{dsE2}). As such, Eq.~(\ref{alt1}) can be used to test the validity of Eq.~(\ref{dsE2}) and, hence, of Eq.~(\ref{vE}) versus Eq.~(\ref{vA}).

\begin{table}[!ht]
    \caption{Alternative formulation of maximum intensity (A-I), three logical blocks of E-PI (E-I, E-II and E-III) and the resulting E-PI and alternative upper limits on maximum velocity.  The alternative estimate assumes that the E-PI assumptions E-II and E-III are valid.}\label{tabEPI}
    \begin{threeparttable}
    \centering    
    \begin{tabular}{m{0.14\textwidth}m{0.28\textwidth}m{0.29\textwidth}m{0.19\textwidth}}
    \hline \hline
     \multicolumn{1}{c}{Atmospheric region}&  \multicolumn{1}{c}{Assumptions}&  \multicolumn{1}{c}{Key relationship}&  \multicolumn{1}{c}{References}\\  
\hline
A-I. Point of maximum wind & The air is in gradient-wind balance; $v_m/r_m \gg f/2$ & $\displaystyle \frac{v_m}{r_m} =  -{\color{black}\frac{1}{\mathcal{C}(1+\zeta)}} T_b \frac{\partial s^*/\partial r}{\partial M/\partial r}$ & Present work\\
    \hline
E-I. Upper atmosphere and the top of  boundary layer   ($z \ge z_b$) & The air is in hydrostatic and gradient-wind balance; surfaces
of constant saturated moist entropy $s^*$ and angular momentum $M$ coincide; $v_m/r_m \gg f/2$ & $\displaystyle \frac{v_m}{r_m} =  -{\color{black}\varepsilon} T_b \frac{\partial s^*/\partial r}{\partial M/\partial r}$ & \citet[][Eqs.~(12), (13)]{em86} \\
    \hline
E-II. Boundary layer near the radius  of maximum wind  ($0 < z \le z_b$)& Horizontal turbulent fluxes of $s^*$ and $M$ are negligible compared to vertical ones; surfaces of constant $s^*$ and $M$ are approximately vertical; turbulent fluxes of $s^*$ and $M$ vanish at $z = z_b$ & $\displaystyle \frac{\partial s^*/\partial r}{\partial M/\partial r} = \frac{\tau_s}{\tau_M} = -\frac{C_k}{C_D}\frac{k_s^* - k_0}{T_s r v_s}$ & \citet[][Eqs.~(32), (33)]{em86},
\citet[][Eqs.~(17), (19), (20)]{emanuel11}\\
\hline
E-III. Air-sea interface near the radius of maximum wind& The upper limit for the air-sea disequilibrium is set by the ambient relative humidity $\mathcal{H}_a$ & 
$k_s^* - k_0 \simeq L_v  (q_s^* - q_0)$, \newline
$q_s^* - q_0 \lesssim (1  - \mathcal{H}_a)q^*_{a}$ & \citet[][p.~3971]{em95}, \citet[][Eq.~(38)]{em89} \\
\hline
E-PI estimate & $T_b \simeq T_s$, $r = r_m$, $v_s(r_m) \simeq v_m$  & $\displaystyle v_m^2 \lesssim \hat{v}_E^2 \equiv {\color{black}\varepsilon} \frac{C_k}{C_D}L_v (1  - \mathcal{H}_a)q^*_{a}$ & \citet[][Eq.~(38) and Table~1]{em89}\\
\hline
Alternative PI estimate & $T_b \simeq T_s$, $r = r_m$, $v_s(r_m) \simeq v_m$ & $\displaystyle v_m^2 \lesssim \hat{v}_A^2 \equiv {\color{black}\frac{1}{\mathcal{C}(1+\zeta)}} \frac{C_k}{C_D}L_v (1  - \mathcal{H}_a)q^*_{a}$ & Present work\\
\hline
\hline
    \end{tabular}
\begin{tablenotes}[para,flushleft]
Notes: $v_m$ and $r_m$ are the maximum velocity and the radius where it is observed; $\tau_s$ and $\tau_M$ are the surface fluxes of, respectively, entropy and angular momentum;  $C_k$ and $C_D$ are exchange coefficients for enthalpy and momentum;  $r$ is local radius; $k_s^*$ is saturated enthalpy at sea surface temperature $T_s$; $k_s^* - k_0 = c_p (T_s - T_0) + L_v (q_s^* - q_0)$, where $c_p$ is the specific heat capacity of air at constant pressure,  $L_v$ is the latent heat of vaporization, $q_s^*$ is the saturated mixing ratio at $T_s$; $v_s$, $k_0$, $q_0$ and $T_0$ are, respectively, the tangential wind speed, enthalpy, water vapor mixing ratio and air temperature at a reference height (usually about 10~m above the sea level); $\mathcal{H}_a$ and $q^*_{a}$ are the relative humidity and saturated mixing ratio at the sea surface temperature in the ambient environment outside the storm core.  For $C_k/C_D \simeq 1$, $\mathcal{H}_a = 0.8$ and $T_b \simeq T_s = 300$~K, we obtain the E-PI upper limit $\hat{v}_E = 60$~m~s$^{-1}$  for $\varepsilon \simeq 0.3$ and the alternative upper limit $\hat{v}_A = 85$~m~s$^{-1}$ for $\mathcal{C} = 1$ (isothermal case).
\end{tablenotes}
\end{threeparttable}
\end{table}

\section{Comparison of conventional and alternative maximum intensities}
\label{isab}

\subsection{Horizontal isothermy}
\label{horis}

Equation~(\ref{dsE2}) (key to E-PI) is valid, i.e., $v_E = v_A$, if 
\begin{equation}\label{con}
\varepsilon = \frac{1}{\mathcal{C} (1+ \zeta)}.
\end{equation}
Since $\zeta$ is proportional to saturated partial pressure of water vapor and increases exponentially with temperature (\ref{Gm}),
Eq.~(\ref{con}) predicts that, for a given $\mathcal{C}$, the value of $\varepsilon$ must decline with increasing temperature.

While the assumption of an isothermal top of boundary layer ($\mathcal{C} = 1$) is not required for deriving $v_E$ (\ref{vE}),
\citet[][cf. his Eqs.~(13), (17) and (26)]{em86} did use this assumption to derive the central surface pressure\footnote{Dr. Steve Garner noted that a factor approximately equal to $1 - \varepsilon(1+\zeta)$ is present in the denominator of \citet{em86}{\textquoteright}s Eq.~(26) for central pressure.  This singularity appeared there because, to derive his Eq.~(26), \citet{em86} simultaneously used  $\varepsilon T_b \partial s^*/\partial r = -\alpha \partial p/\partial r$ and a version of our Eq.~(\ref{sr}) for unsaturated isothermal air,  \citet{em86}{\textquoteright}s Eqs.~(21) and (25), respectively,  as if they were independent constraints. Since, in view of Eq.~(\ref{diff}), they are not, \citet{em86}{\textquoteright}s Eq.~(26) is an identity. Hypercanes introduced based on this equation are a misconception, see appendix~\ref{hyper} for details.}. Temperature $T_b$ of the top of the boundary layer was taken {\textquotedblleft}to be constant as is generally observed \citep[e.g., see][]{frank77}{\textquotedblright} \citep[][p.~588]{em86}. \citet[][p.~2245]{emanuel11} also assumed $T_b = \rm const$. This apparently plausible assumption deserves a special consideration.

\begin{figure*}[tbp]
\begin{minipage}[p]{0.85\textwidth}
\centering\includegraphics[width=1\textwidth,angle=0,clip]{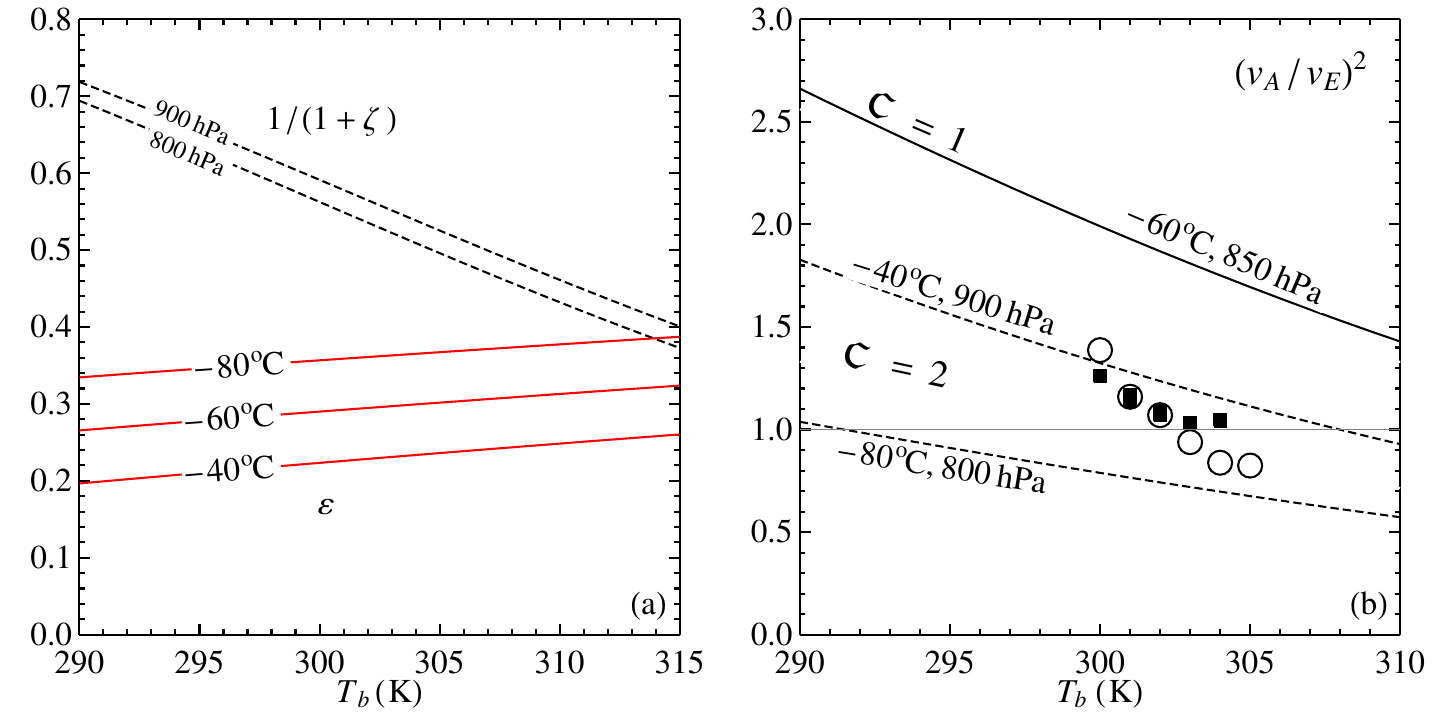}
\end{minipage}
\caption{Parameters $\varepsilon$ versus $1/(1+\zeta)$ (a)  and $(v_A/v_E)^2$ (b) as dependent on temperature $T$. In (a), the $\varepsilon \equiv (T_b - T_o)/T_b$, with $T_b = T$ curves correspond to different outflow temperatures $T_o$;  the $1/(1+\zeta)$ curves correspond to $p_d$ values of  $800$ and $900$~hPa, see Eq.~(\ref{Gm}). In (b), the solid curve and the dashed curves correspond to $\mathcal{C} = 1$ and $\mathcal{C} = 2$, respectively, in Eq.~(\ref{diff}); $T_o$ and $p_d$ used to calculate $\varepsilon$ and $\zeta$ are shown at the curves. Open circles and solid squares correspond to $(v_L/v_E)^2$ from, respectively, Fig.~2a and Fig.~11c of \citet{Li20c}, where $v_L$ is the maximum surface wind speed derived from their model and $v_E$ is the maximum intensity calculated by \citet{Li20c} from Eq.~(\ref{vE}) with sea surface temperature used instead of $T_b$ in $\varepsilon$.}
\label{fig1}
\end{figure*}

With $\mathcal{C} = 1$, Eq.~(\ref{diff}) is not satisfied under common atmospheric conditions (Fig.~\ref{fig1}a). The maximum Carnot efficiency estimated from the temperatures observed in the outflow and at the top of the boundary layer  is $\varepsilon= 0.35$ \citep{demaria94}. Assuming that $T_b$ does not usually exceed $303$~K ($30${\textcelsius}), the minimum value of $1/(1+\zeta)=0.5$ is $1.4$-fold larger. It corresponds to the largest $\gamma_d^* \simeq 0.05$ for $T_b = 303$~K  and $p_d \simeq p = 800$~hPa. The  partial pressure $p_v^*$ of saturated water vapor  and, hence, $\gamma_d^*$ depend exponentially on air temperature. The realistic temperatures at the top of the boundary layer  are commonly significantly lower than $303$~K. Thus, the discrepancy between $v_E$ and $v_A$ should be commonly significantly higher (Fig.~\ref{fig1}a). The mismatch diminishes when $\mathcal{C} > 1$, i.e., when $T_b$ increases towards the storm center.

Since $1/(1 + \zeta)$  declines with $T$, the discrepancy between $v_E$ and $v_A$ diminishes with increasing temperature (Fig.~\ref{fig1}). 
This explains why, beyond a certain temperature, superintensity $v_A/v_E > 1$ becomes impossible and changes to what \citet{Li20c}, who established this pattern in a numerical modelling study, termed {\textquotedblleft}sub-MPI  intensity{\textquotedblright}. The particular temperature at which this shift occurs depends on the value of $\mathcal{C}$. Figure~\ref{fig1}b shows that $(v_L/v_E)^2$, where $v_L$ is maximum intensity produced by the model of \citet{Li20c}, declines with growing temperature  faster than does $(v_A/v_E)^2 = 1/[\varepsilon \mathcal{C}(1 + \zeta)]$, Eq.~(\ref{diff}), if the outflow temperature $T_o$, $\mathcal{C}$ and $p$ (the latter enters the definition of $\zeta$) are assumed to be temperature-independent. This indicates that $\mathcal{C}$ should increase, and/or $p$ and/or $T_o$ should decrease, at higher temperatures,  propositions that could be tested in future studies\footnote{Note that \citet{Li20c} applied Eq.~(\ref{vE}) to surface wind, so $p$ and $T$ for their data shown in our Fig.~\ref{fig1}b refer to surface pressure and temperature at the point of maximum wind. Since they observed an increased storm intensity at higher surface temperatures, a decrease of surface pressure is quite plausible.}.

\subsection{Supergradient wind and Hurricane Isabel 2003}
\label{hurisab}

\begin{figure*}[tbp]
\begin{minipage}[p]{0.85\textwidth}
\centering\includegraphics[width=0.5\textwidth,angle=0,clip]{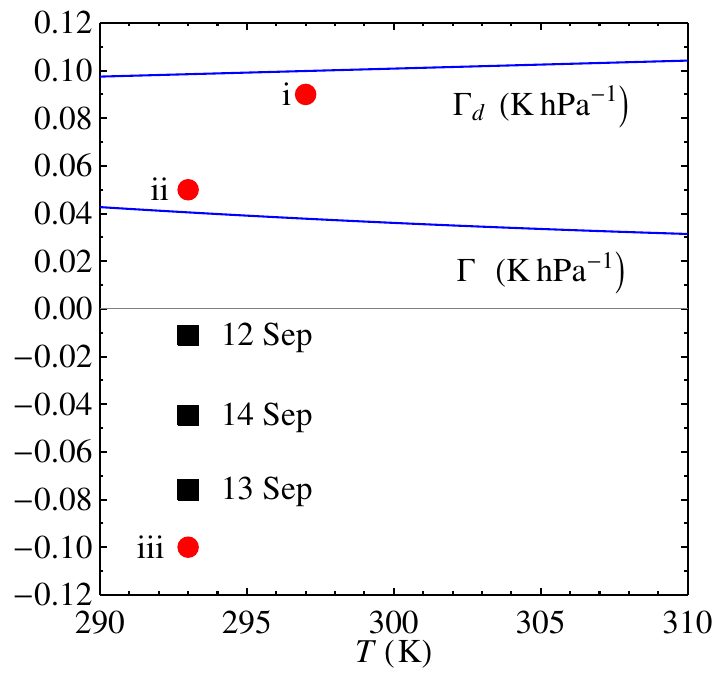}
\end{minipage}
\caption{Moist $\Gamma$  and dry $\Gamma_d$ adiabatic temperature gradients as dependent on temperature. They are calculated for $p = 850$~hPa, see Eq.~(\ref{Gm}). The circles indicate the {\it mean} temperature gradients (K~hPa$^{-1}$) observed in Hurricane Isabel 2003 on September 13 between the eyewall and the outer core at the surface (i) and at the top of the boundary layer  (ii) and between the eye and the eyewall at the top of the boundary layer  (iii); the squares indicate the {\it local} temperature gradients (K~hPa$^{-1}$) at the point of maximum wind calculated from Eq.~(\ref{sr1}) and the data of Table~\ref{tabisab} for Hurricane Isabel 2003 on September 12, 13 and 14.  Note that negative values correspond to temperature increasing towards the storm center. See section~\ref{hurisab} for calculation details.
}
\label{fig2}
\end{figure*}

In the general case, instead of the gradient-wind balance (\ref{gwb}), we can write
\begin{equation}\label{gwbg}
\alpha\frac{\partial p}{\partial r} \equiv \mathcal{B} \frac{v^2}{r},
\end{equation}
where $\mathcal{B}$ defines the degree to which the flow is radially unbalanced: $\mathcal{B} < 1$ for the supergradient flow when the outward-directed centrifugal force is larger than the inward-pulling pressure gradient. For example, in the numerical experiment of \citet[][their Fig.~8]{bryan09b}, $\mathcal{B} \simeq 0.5$ at the point of maximum wind, but $\mathcal{B} \simeq 1$ at the radius of maximum wind at the surface (see Fig.~8 of \citet{bryan09b} and appendix~\ref{bryan}).  (Notably, if the gradient-wind imbalance is indeed minimal at the surface, it could not explain superintensity of surface winds in the simulations of \citet{Li20c}.)

If, under supergradient conditions, $v/r \gg  \partial v/\partial r \simeq 0$ at the point of maximum wind, we have $\partial M/\partial r \simeq v + fr \simeq v$ (assuming $v/r \gg f$) and Eq.~(\ref{sr}) can be written as
\begin{equation}\label{sr1}
\frac{1}{(1+\zeta)}\frac{T_b }{\mathcal{B} \mathcal{C}} \frac{\partial s^*}{\partial r} = - \frac{v_m^2}{r_m}.
\end{equation}
In comparison, Eq.~(\ref{dsE2}) at the point of maximum wind becomes \citep[see, e.g.,][their Eq.~(5)]{bell08}:
\begin{equation}\label{cf}
\varepsilon T_b \frac{\partial s^*}{\partial r} = - \frac{v_E^2}{r_m}.
\end{equation}

Comparison of Eq.~(\ref{sr1}) with Eq.~(\ref{cf}) reveals that the flow being supergradient ($\mathcal{B}<1$)  and air temperature declining towards the storm center ($\mathcal{C} < 1$)  cause $v_E$ to underestimate maximum velocity $v_m$ (\ref{sr1})  more than it does in the radially balanced isothermal case ($\mathcal{B}=1$, $\mathcal{C}=1$). Conversely, for E-PI{\textquoteright}s Eq.~(\ref{cf}) to be consistent with observations for a radially balanced  ($\mathcal{B} = 1$) or supergradient ($\mathcal{B} \leq 1$) flow, the temperature at the point of maximum wind must increase towards the hurricane center ($\mathcal{C} > 1$).

When, as is the case in the stronger storms,  the pressure gradient is sufficiently steep and the radial motion sufficiently rapid, the radial expansion of air is accompanied by a drop of temperature. In the well-studied  Hurricane Isabel 2003 the surface air cooled by about 4~K while moving from the outer core (150-250~km) to the eyewall (40-50~km) \citep[][their~Fig.~4c]{montgomery06}. Over the same distance, the surface pressure fell by less than 50~hPa (from less than 1013~hPa to 960~hPa) \citep[][their~Fig.~4]{aberson06}. (Air pressure at the outermost closed isobar $\sim 465$~km from the center was 1013~hPa, hence at 150-250~km from the center it should have been smaller.)   With $\Delta p \simeq -45$~hPa and $\Delta T \simeq -4$~K, at $T \simeq 300$~K  and $p \simeq 10^3 $~hPa, the horizontal temperature gradient at the surface $\Delta T/\Delta p = 0.09$~K~hPa$^{-1}$ approaches the dry adiabatic gradient $\Gamma_d = \mu T/p = 0.1$~K~hPa$^{-1}$ (see Fig.~\ref{fig2}). For their numerical simulation of an intense cyclone, \citet{kepert16} reported a drop of $2$~K in surface air temperature from $100$~km to $50$~km from the center, which is a temperature gradient comparable to Hurricane Isabel{\textquoteright}s.
In contrast, earlier observational studies concluded that the temperature drop within $100$~km from the storm center was on average at least several times lower \citep[][their Fig.~1]{barnes01} or nearly non-existent \citep[][their Fig.~3a]{cione00}, while at larger distances the adiabatic expansion was insufficient to explain the observed horizontal temperature differences \citep{cione00,barnes01}.
	
At the top of the boundary layer  the radial flow is weaker than it is on the surface, and the mean horizontal temperature gradient is smaller \citep[][their Fig.~4b,c]{montgomery06}. At the level of maximum wind $z_m = 1$~km in Hurricane Isabel 2003 the temperature difference between the eyewall and the outer core  was $\Delta T \simeq -2$~K \citep[][their Fig.~4c]{montgomery06}. Assuming that the pressure difference at this level is about $0.9$ of its value at the surface, see  Eq.~(\ref{ptr}), $\Delta p \simeq -0.9 \times 45$~hPa, we have $\Delta T/\Delta p = 0.05$~K~hPa$^{-1}$. The mean horizontal temperature gradient between the outer core and the eyewall at the top of the boundary layer  approaches the moist adiabatic gradient  $\Gamma = 0.04$~K~hPa$^{-1}$ for $T =  293$~K and $p =  850$~hPa (Fig.~\ref{fig2}). 

In the eye, the surface heat fluxes and the descending air motion work to elevate the air temperature above that at the eyewall 
\citep[see, e.g.,][their Fig.~1]{barnes01}. The air temperature in the eye rises towards the storm center and $\partial T/\partial r < 0$. For Hurricane Isabel 2003, with pressure and temperature differences at $z_m = 1$~km between the eye and the eyewall $\Delta p \simeq -30$~hPa 
\citep[][their Fig.~4]{aberson06} and $\Delta T = 3$~K \citep[][their Fig.~4c]{montgomery06}, for $p = 850$~hPa and $T =293$~K we have $\Delta T/\Delta p = -0.1$~K~hPa$^{-1}~\simeq -2.5 \Gamma$ (Fig.~\ref{fig2}).

That the horizontal temperature gradient changes its sign somewhere in the eyewall suggests that $\partial T/\partial r = 0$ at the point of maximum wind is a plausible assumption (see section~\ref{dTdr}). However, the magnitudes of horizontal temperature gradients on both sides of the eyewall are large enough to significantly impact the maximum velocity  estimates (Fig.~\ref{fig2}). 

For example, if at the point of maximum wind the horizontal temperature gradient were close to $\Gamma$ (as it was on average between the eyewall and the outer core in Hurricane Isabel 2003), then $\mathcal{C} \to 0$ and $\mathcal{B}\mathcal{C}(1+\zeta)\varepsilon \to 0$. In this case E-PI{\textquoteright}s $v_E$ would formally infinitely {\it underestimate} $v_m$ (\ref{sr1}). Physically, this limit corresponds to the situation when the moist adiabat is locally horizontal, $\partial s^*/\partial r \to  0$, such that the dependence between saturated moist entropy and maximum velocity vanishes, see Eq.~(\ref{sr1}).

If, on the other hand, at the point of maximum wind the horizontal temperature gradient were equal to $-2.5\Gamma$ (as it was on average between the eye and the eyewall in Hurricane Isabel 2003), then  $\mathcal{C} = 3.5$. In this case, E-PI{\textquoteright}s $v_E^2$ would {\it overestimate} $v_m^2$ by $\mathcal{B}\mathcal{C}(1+\zeta)\varepsilon = 1.4$-fold for a balanced flow ($\mathcal{B} = 1$). With $\mathcal{B} = 0.8$ as discussed above, the overestimate reduces to $1.1$ (about $10\%$).

For a known $\mathcal{B}$, the value of $\mathcal{C}$ can be derived from the observed values of variables entering Eq.~(\ref{sr1}).
The data for Hurricane Isabel 2003 suggest that at the point of maximum wind the air temperature increases towards the center $\mathcal{C} > 1$, but not enough to bring E-PI in agreement with observations: on September 12 and 14, E-PI{\textquoteright}s  Eq.~(\ref{cf}) underestimates the observed squared maximum velocity $v_m^2$ by about $50\%$ and $30\%$, respectively (Table~\ref{tabisab}).

\begin{table}[!t]
    \caption{Parameters of Eqs.~(\ref{sr1}) and~(\ref{cf}) estimated from observations for Hurricane Isabel 2003.}\label{tabisab}
    \begin{threeparttable}
    \centering    
    \begin{tabular}{ccccccccccc}
    \hline \hline
    Date&$r_m$, km&$T_o$, {\textcelsius}&$\theta_e$, K&$\partial \theta_e/\partial r$,~K~km$^{-1}$&$v_m$, m~s$^{-1}$&$v_E$, m~s$^{-1}$&$\varepsilon$&$\mathcal{B}$&$(v_m/v_E)^2$ & $\mathcal{C}$\\
    \hline
    12 September&$25$&$-65$&$360$&$-0.5\hphantom{3}$&$80$&$54$&$0.29$&$0.95$&$2.2$ & $1.1$\\ 
    13 September&$45$&$-58$&$357$&$-0.6\hphantom{3}$&$76$&$74$&$0.27$&$0.72$&$1.1$ & $3.2$\\ 
    14 September &$50$&$-56$&$357$&$-0.35$&$74$&$61$&$0.26$&$0.85$&$1.5$ & $2.2$\\ 
    \hline        
    \end{tabular}
\begin{tablenotes}[para,flushleft]
Observed values of  the radius of maximum wind $r_m$, outflow temperature $T_o$,  maximum velocity $v_m$ and  $\partial \theta_e/\partial r$ are taken from, respectively, the first and the third columns of Table~2 of \citet{bell08}, and equivalent potential temperature $\theta_e$ from their Fig.~5  at $r = r_m$ and $z = 1$~km; $\mathcal{B}$ is calculated from Eq.~(\ref{B}) using data from \citet{bryan09b}{\textquoteright}s Table 1 (see appendix~\ref{bryan}); $\mathcal{C}$ is calculated from Eq.~(\ref{sr1}). At the top of the boundary layer,  temperature $T_b = 293~K (20${\textcelsius}) corresponding to $z_b = 1$~km is assumed for all the three days based on Fig.~4c of \citet{montgomery06}, $\zeta = L\gamma_d^*/RT_b = 0.49$ for $p_d = 850$~hPa and $p_v^*=23$~hPa. The values of $v_E$ are E-PI estimates of maximum velocity obtained from Eq.~(\ref{cf}), where $\partial s^*/\partial r = (c_p/\theta_e) \partial \theta_e/\partial r$, $c_p = 1$~kJ~kg$^{-1}$~K$^{-1}$  \citep[see][their Eq.~(A2)]{montgomery06}.
\end{tablenotes}
\end{threeparttable}
 \end{table}

The closest agreement is observed on September 13, when $\mathcal{C}$ is the largest (Fig.~\ref{fig2}). Given that the flow at this date was supergradient with $\mathcal{B} \simeq 0.8$ \citep{bell08}, this agreement does not indicate that the storm is in thermal wind balance \citep[cf. ][p.~1345]{montgomery06}. Rather, it suggests that the large value of $\mathcal{C} = 2.9$ nearly compensated the underestimate  that would have otherwise resulted from $\mathcal{B} < 1$.  The underestimate is greatest on September 12, when the local temperature gradient is closest to zero (Fig.~\ref{fig2}) and $\mathcal{C}$ is close to unity (Table~\ref{tabisab}). The data in Table~\ref{tabisab} indicate that superintensity cannot be explained by supergradient winds alone (see appendix~\ref{bryan}),  because maximum superintensity observed on September 12 corresponds to minimal gradient-wind imbalance ($\mathcal{B} = 0.95$).

\section{Discussion}
\label{concl}

\subsection{The nature and magnitude of the horizontal temperature gradient}
\label{dTdr}

The alternative expression for maximum potential intenstity, Eq.~(\ref{vA}), shows how the magnitude of the radial temperature gradient in the cyclone core can be used to assess E-PI{\textquoteright}s validity. Until now it has not received much attention in such assessments \citep[e.g.,][]{montgomery06,bryan09b,emanuel11,wang20,wang21}. There were some discussions of possible changes of outflow temperature \citep[e.g.,][]{emanuel11,montgomery2019,ms20}, but regarding the characteristic magnitude, or even sign, of $\partial T/\partial r$ at the point of maximum wind, the literature is offering no clues. Here we provide a few initial considerations.

The energy budget of an air parcel at the sea surface that does not contain liquid water and moves horizontally with radial velocity $u$ and total velocity $V$, can be written as
\begin{equation}\label{u}
c_p \frac{dT}{dt} = \alpha_d \frac{dp}{dt} + \frac{\delta Q}{dt} \simeq u \alpha \frac{\partial p}{\partial r} \left[ 1 +  \left(\frac{\partial p}{\partial r}\right)^{-1} \frac{\rho C_k c_p}{z_m}(T_s - T_0) \frac{V}{u}\right].
\end{equation}
Here we neglected horizontal diffusion, as did \citet{em86}, and assumed that all heat $\delta Q/dt$ (W~kg$^{-1}$) that the air parcel receives, comes from local surface flux of sensible heat $J_S=\rho C_k c_p(T_s - T_0)V$ (W~m$^{-2}$) (see Table~\ref{tabEPI} for other notations). We assumed that all this surface flux of sensible heat is absorbed below a certain level $z_J$, and, in the second equality of Eq.~(\ref{u}), approximated  the volume-specific heat input $\rho \delta Q/dt$ (W~m$^{-3}$) by 
\begin{equation}\label{dQ}
\rho \frac{\delta Q}{dt} = \frac{J_S}{z_J} = \frac{\rho C_k c_p}{z_m}(T_s - T_0) V,
\end{equation}
under the assumption that $z_J = z_m$, where $z_m$ is the level of maximum wind. With common values of $c_p = 1$~kJ~kg$^{-1}$~K$^{-1}$, $C_k \simeq 10^{-3}$, $\rho = 1$~kg~m$^{-3}$,  and with $\partial p/\partial r = 0.03$~Pa~m$^{-1}$, $T_s - T_0 = 3$~K  and $z_m = 1$~km as in Hurricane Isabel 2003 \citep[][]{montgomery06}, the factor in front of the fraction $V/u$  inside the square brackets in Eq.~(\ref{u}) is numerically equal to $0.1$. Since in Hurricane Isabel 2003 at the surface $V/u \simeq -2$ both in the outer core and in the eyewall \citep[][their Fig.~4a,b]{montgomery06}, Eq.~(\ref{u}) predicts a $20\%$ reduction from dry adiabaticity for the horizontal temperature gradient of the surface flow. The actual reduction is only $10\%$ (see Fig.~\ref{fig2}, point i). This suggests that $z_J = z_m$ could be an underestimate or that some minor additional cooling is provided by other mechanisms (e.g., by subcloud evaporation not accounted for in Eq.~(\ref{u})). Equation~(\ref{u})  shows that adiabaticity of the horizontal temperature gradient is controlled by radial velocity  rather than being determined by the horizontal pressure gradient alone.

Equation~(\ref{u}) also indicates that in intense cyclones the air-sea temperature disequilibrium in the eyewall can be largely a product of cyclone{\textquoteright}s secondary circulation. As the air moves over an isothermal oceanic surface and cools due to expansion, the air-sea temperature disequilibrium increases. In Hurricane Isabel 2003, the air cooled by $4$~K as it moved from the outer core to the eyewall -- this compares well to the estimated $T_s - T_0 \simeq 3$~K in the inner core.  The disequilibrium is by $1$~K smaller than the adiabatic cooling of the air because the oceanic surface was also $1$~K colder in the inner core than in the outer core \citep[$27.5^{\rm o}$C versus $28.5^{\rm o}$C,][]{montgomery06}. This cooling of the oceanic surface in the eyewall can be attributed to different causes like turbulent mixing of the upper oceanic level by hurricane winds \citep{montgomery06} or to a smaller flux of solar radiation in the eyewall as compared to the storm{\textquoteright}s outskirts \citep{zhou2017}. Furthermore, if the surface air in the eyewall has cooled appreciably as compared to its ambient environment, it must have a high relative humidity, up to saturation. Thus, saturated air near the radius of maximum wind is also a product of cyclone{\textquoteright}s activity. Once the air reaches saturation, it cannot easily cool further as there appears an additional source of heat (latent heat) not reflected in the right-hand part of Eq.~(\ref{u}).

A stronger air-sea disequilibrium due to the drop of air temperature at the radius of maximum wind in the stronger storms is visible in models.  For example, in Fig.~9c of \citet{wang20} in the strongest tropical cyclone (maximum wind speed over 90~m~s$^{-1}$) there is a pronounced local maximum of air-sea disequilibrium in the vicinity of maximum wind. In contrast, the weaker cyclones (maximum wind speeds about 50~m~s$^{-1}$) in Figs.~9a and 9b of \citet{wang20}, as well as all cyclones in Fig.~5 of \citet{wang21} (maximum wind speeds below $65$~m~s$^{-1}$), display a more monotonic decline of the air-sea disequilibrium from the outer environment towards the center. (Notably, while reporting these distinct patterns of air-sea disequilibrium, \citet{wang20} and \citet{wang21} did not analyze the magnitude  of either radial velocity or radial gradient of air temperature.) For observed cyclones, the horizontal flow was still isothermal in Hurricane Earl 2010 (maximum wind speed 64~m~s$^{-1}$)  but approached dry adiabatic in Hurricane Isabel 2003 (maximum wind speed 75~m~s$^{-1}$) \citep{smith13,bell08}. This suggests, as a hypothesis for further studies, that adiabaticity is approached in very intense cyclones only.

Between the eyewall and the eye the radial velocity changes its sign, so at a certain point, specifically when $V/u \simeq -10$ for the case of Hurricane Isabel 2003, the temperature gradient at the surface must turn to zero. This happens somewhere {\it within the eyewall}, i.e., close to the radius of maximum wind. If the surface air at the point where $\partial T_s/\partial r = 0$ is saturated, as it approximately was in the eyewall in Hurricane Isabel 2003 \citep[][their Fig.~4d]{montgomery06} or Hurricane Earl 2010 \citep[][their Table~1]{smith13}, the radial temperature gradient at the level of maximum wind $z = z_m$,  and the corresponding value of $\mathcal{C}$, will be approximately the same as they are at the surface (see Eq.~\eqref{Cb}). These arguments justify the plausibility of horizontal isothermy ($\mathcal{C} = 1$) at the point of maximum wind. Using the gradient-wind and hydrostatic balances and a typical tangential wind profile of a tropical cyclone, \citet[][his Fig.~1b]{smith2007} calculated that, at the radius of maximum wind, $\partial T/\partial r$ is positive at the surface, negative in the free troposphere, and changes sign ($\partial T/\partial r = 0$) at $z \sim 1$~km, i.e., close to a typical altitude of maximum wind. It should be noted, however, that in \citet[][]{smith2007}{\textquoteright}s calculations there was no temperature increase in the eye as compared to the radius of maximum wind \citep[cf.][their Fig.~1]{barnes01}.

As the surface air moves further towards the center beyond the point where $\partial T_s/\partial r = 0$, it begins to warm. As it is warmed by the surface heat flux, the temperature disequilibrium across the air-sea interface should diminish. Thus, an increase in $\mathcal{C}$ that reduces $v_A$ is accompanied by a decrease in $T_s - T_0$ that should reduce $v_A$ even further, see Eq.~(\ref{vA}).  The complex interplay of these influences, and their profound dependence on the storm{\textquoteright}s dynamic structure and the relation between the primary and secondary circulations ($v/u$),  may help explain why maximum velocities, for a given environment, strongly depend on the characteristics of the initial vortex \citep{tao20}. Generally, $\mathcal{C} > 1$ at the point of maximum wind can result from the surface air warming, and/or from the surface relative humidity increasing, towards the center  (see Eqs.~\eqref{Cb} and \eqref{iCb}).

\subsection{External and internal parameters in the maximum potential intensity formulations}
\label{ext}

While in our alternative formulation $v_A$ (\ref{vA}) the value of $\mathcal{C}$ is determined by the internal structure of the cyclone, E-PI has been characterized as a closed theory that allows the estimation of storm{\textquoteright}s maximum speed from environmental parameters alone (compare $\hat{v}_A$ and $\hat{v}_E$ in Table~\ref{tabEPI}). In this interpretation, the outflow temperature $T_o$ (which corresponds to the point where $v = 0$ and is, strictly speaking, a property of the cyclonic flow itself rather than of its environment) is assumed to approximately coincide with the temperature at which an air parcel saturated at ambient surface temperature and raised moist adiabatically in an environmental sounding becomes neutrally buoyant \citep[e.g.,][]{ro87,wang2014}.

The environmental soundings assumed to be representative of the outflow location  are commonly measured at a distance of $300$-$700$~km from the storm center \citep[e.g.,][]{montgomery06}. At these radii the environment experiences a strong influence of the cyclone. For example, in North Atlantic hurricanes at a distance of $400$~km from the center the column water vapor content and precipitation rate are, respectively, $15\%$  and two times higher than they are in hurricane absence \citep[][Fig.~4a,g]{ar17}. In Hurricane Isabel 2003 on September 12,
mean precipitation rate at the radius of the outermost closed isobar ($\sim 400$~km) was ten times the local climatological mean in hurricane absence \citep[][Fig.~11]{ar17}. Tropical storms can also perturb the tropopause temperature by up to 3~K \citep{ratnam2016}.

These empirical findings have two implications. First, calculating $\hat{v}_E$ (Table~\ref{tabEPI}) for an arbitrary environment may not be very informative: if the cyclone modifies its outflow region, there should exist {\textquotedblleft}unmodified{\textquotedblright} environments where cyclonic outflows may never happen. On the other hand, these observations indicate that the cyclonic flow emanating to the free troposphere from the point of maximum wind with a given $\mathcal{C}$ could, in principle, transform the downstream environment to such a degree that the equality $v_A = v_E$ will hold. In this case, Eq.~(\ref{diff}) will define the outflow temperature $T_o$ as follows:
\begin{equation}\label{To}
T_o = T_b \left[1 - \frac{1}{\mathcal{C} (1+ \zeta)} \right].
\end{equation}
For Lilly{\textquoteright}s analytical model, which is closely related to E-PI, it was established that the outflow temperature
(which Lilly himself believed was environmentally prescribed) is controlled by the interior flow of the cyclone \citep{tao20b}.
Equations~(\ref{con}) and (\ref{To}) are probably more transparent than, but have a similar physical meaning as, \citet{tao20b}{\textquoteright}s Eq.~(2), although demonstrating the equivalence of different maximum potential intensity formulations may require considerable space \citep[cf.][]{mpi4-jas}.

For $\mathcal{C} \simeq 1$ and a realistic $T_b$, Eq.~(\ref{To}) predicts that the outflow temperature should be about one half of $T_b$, i.e., around $150$~K. Such temperatures are well below the tropopause temperature and cannot be realized. When the flow approaches the tropopause, it ceases to be adiabatic thus violating one of the key E-PI{\textquoteright}s assumptions for the free troposphere. For the flow to be adiabatic, as the first equality in Eq.~(\ref{u}) reminds, the external heat input must  be significantly smaller than the change of internal energy due to expansion.  But closer to the tropopause far from the cyclone core the horizontal pressure gradient vanishes, while the vertical motion cannot be adiabatic due to the fact that the stratospheric warming increases rapidly as $T_o$ diminishes below the tropopause temperature $T_t$ \citep[e.g.,][their Eq.~(1)]{wang2014}.  (The stratosphere constrains the height of the outflow due to the fact that the adiabatically ascending air becomes negatively buoyant. Non-zero buoyancy implies a radical increase of the temperature gradient between the ascending air and its environment -- compared to the tropospheric motions assumed in E-PI to be neutrally buoyant. This extra temperature gradient causes warming that breaks the adiabaticity of the flow especially as the vertical velocity (responsible for expansion) diminishes.)

\citet{mpi4-jas} showed that conventional E-PI at the point of maximum wind corresponds to an infinitely narrow thermodynamic cycle, where the finite changes of temperature and pressure are adiabatic and where total work in the free troposphere is zero. Work of this cycle is equal to work at  the top of the boundary layer $z_m = z_b$ and to heat input (that also occurs at $z_m = z_b$) multiplied by Carnot efficiency. The cycle being infinitely narrow, its efficiency does not depend on the infinitely small change of temperature and/or relative humidity at the boundary layer (that is why E-PI did not require any assumptions about horizontal temperature gradient). But it does depend on the assumed adiabaticity of the finite parts of the cycle. The stratospheric warming violates this E-PI assumption and makes the formula for $v_E$ (\ref{vE}) invalid.

The interpretation of E-PI as a cycle with zero work in the free troposphere helps explain this fact.  For brevity, we consider a saturated isothermal boundary layer \citep[an isothermal path  $\rm B'b$ in Fig.~1b of][]{makarieva20}, but extension for an arbitrary temperature gradient and unsaturated conditions is straightforward. At the boundary layer (path  $\rm B'b$) we have heat input $Q_{\rm in} = L_vdq^* - \alpha_d dp$ and work $A = - \alpha_d dp = \varepsilon Q_{\rm in}$. In the free troposphere \citep[path  $\rm bcC'B'$ in Fig.~1b of][]{makarieva20} from $\delta Q = c_p dT - \alpha_d dp + L_v dq^*$  \eqref{Tds2ss} we find that, as far as the integrals of $c_p dT$ and $\alpha_d dp$ are zero,  the cycle{\textquoteright}s heat output equals $Q_{\rm out} = -L_vdq^* < 0$. Since $A = Q_{\rm in} + Q_{\rm out}$, we have  $-\alpha_d dp =[\varepsilon/(1 - \varepsilon)] L_vdq^*$. On the other hand, for an isothermal saturated case from the Clausius-Clapeyron law we have  $dq^*/q^* = -dp/p_d$  \citep[see, e.g., Eq.~(3) of \citet{ar17} or Eq.~(9) of][]{tao20b}. Combining this\footnote{To our knowledge, the first mention of the discrepancy between the E-PI formulation and the alternative one stemming from the definition of $dq$, as illustrated by Eqs.~(\ref{con}) and (\ref{To}), was by \citet[][their Eqs.~(40) and~(43)]{makarieva18b}.} with the previous expression  and noting that $\zeta \equiv L_v q^*/(\alpha_d p_d)$, see Eq.~(\ref{Gm}), we obtain Eq.~(\ref{con}) with $\mathcal{C} = 1$. In other words, the difference between $v_A$ and $v_E$ results from E-PI relating $L_v dq^*$ to $\alpha_d dp$ via Carnot efficiency $\varepsilon$, while the same magnitudes are distinctly related by the Clausius-Clapeyron law.

When the adiabaticity in the free troposphere is perturbed by stratospheric warming, such that heat output is no longer $-L_vdq^*$, the relationship between work and heat input is perturbed as well\footnote{There are certainly other processes that can perturb adiabaticity like dry air ventilation into the eyewall, but they diminish the cyclone intensity as compared to E-PI \citep[e.g.,][]{tang10} as opposed to the stratospheric warming that can cause superintensity.}, and the E-PI formulation no longer holds. The implication is that in those cases when $\mathcal{C}$ is sufficiently small, while the observed outflow temperature $T_o$ approaches the tropopause temperature, $T_o \simeq T_t$,  $v_E$ will underestimate $v_m$ due to the violation of adiabaticity. Such {\textquotedblleft}superintensity{\textquotedblright} (unexplained by supergradient winds, see  appendix~\ref{bryan})  was found by \citet{wang2014} who forced the tropopause temperature to be constant.  When, on the other hand, $T_o \ll T_t$ such that Eq.~(\ref{To}) could hold, the potential of E-PI  to predict $v_m$ from environmental parameters (e.g., from $T_t$) is diminished.

\subsection{Why $v_E$ underestimates, while $\hat{v}_E$ overestimates, maximum winds}
\label{under}

If horizontal isothermy is a common condition under which $v_E$ (\ref{vE}) underestimates $v_m$, one has to explain why in most cases  the maximum wind speeds observed in real cyclones are well below the environmental version of E-PI, $\hat{v}_E$ (Table~\ref{tabEPI}). Since the underestimate $v_E < v_m$ results from E-PI assumptions pertaining to the free troposphere (block E-I in Table~\ref{tabEPI}), the overestimate
$\hat{v}_E > v_m$ indicates that a certain overcompensation should occur in the assumptions pertaining to the remaining two E-PI blocks,  the boundary layer interior and the air-sea interface (Table~\ref{tabEPI}, blocks E-II and E-III).

One compensating overestimate should result from E-PI{\textquoteright}s assumptions concerning the disequlibrium $\Delta k = k_s^* - k_0 = c_p \Delta T + L_v \Delta q$ at the air-sea interface at the radius of maximum wind.  Since the local enthalpy difference $\Delta k$ is unknown, E-PI limits it from above by assuming that the local difference in mixing ratios $\Delta q$ is less than the water vapor deficit $(1 - \mathcal{H}_a) q_{a}^*$ in the ambient environment (Table~\ref{tabEPI}, block E-III). However, as \citet{em86} and \citet{emanuel11} pointed out, in reality $\Delta k$ tends to decline from the outer core towards the storm center. Indeed, if the radial inflow is sufficiently slow, as it is in the weaker storms, the surface air can remain in approximate thermal equilibrium with the oceanic surface.  In his original evaluations of E-PI \citet[][p.~591]{em86} assumed $\Delta T = 0$. On the other hand, evaporation into the air parcels that are spiraling inward increases the relative humidity and diminishes $\Delta q$. As a result, in the weaker cyclones the actual $\Delta k$ at the radius of maximum wind can be much lower than its ambient constraint $(1 - \mathcal{H}_a) q_{a}^*$.  This would overcompensate the underestimate of the observed $v_m$ at the top of the boundary layer  by E-PI{\textquoteright}s Eq.~(\ref{cf}) and, provided the assumptions in block E-II hold, explain why in most cases the E-PI upper limit $\hat{v}_E$ (Table~\ref{tabEPI}) goes above the observed maximum velocities.

In the stronger storms, as we discussed, the air streams so quickly towards the center that it cools significantly compared
to the isothermal oceanic surface it moves above. As pointed out by \citet{CampMontgomery01} and \citet{montgomery06}, this cooling tends to offset the increase in relative humidity,  such that the mixing ratio $q$ does not considerably grow, and $\Delta q$ does not diminish significantly, towards the center.  In this case the E-PI{\textquoteright}s assumption, $\Delta q \simeq (1 - \mathcal{H}_a) q_{a}^*$, becomes valid. No overcompensation occurs in the third block of E-PI. As a result, in the strongest storms the underestimate stemming from the first block of E-PI manifests itself and $\hat{v}_E < v_m$.

A distinct type of compensation can occur between the temperature gradient and the supergradient wind (coefficients $\mathcal C$ and $\mathcal B$ in Eq.~(\ref{sr1})).  We have shown that supergradient winds in the formulation of \citet{bryan09b} are not sufficient to explain the mismatch between E-PI and actual maximum velocities in either numerical simulations or real cyclones (Table~\ref{tabisab} and  appendix~\ref{bryan}). Our new formulation explains that, despite E-PI is assumed to underpredict supergradient winds ($\mathcal B < 1$), it can nevertheless match the observations when $\mathcal C > 1$, i.e., when the winds are supergradient but the temperature at the point of maximum wind rises towards the center. This appears to be the case in Hurricane Isabel 2003 on September 13 (Table~\ref{tabisab}). In other words, the question  {\it why E-PI in most cases overestimates observed intensities}, is directly relevant to {\it why it sometimes underestimates them}, i.e., to the superintensity problem. 

Finally, there is a major uncertainty pertaining to the second block of E-PI, i.e., to the transition from the 
volume to surface fluxes of entropy and momentum, Eq.~(\ref{tau}). This transition, while key to both conventional E-PI, to its recent modification for the surface winds \citep{re19},  and to any local formulation of maximum intensity based on Eq.~(\ref{tau}), including $v_A$ (\ref{vA}), has not been rigorously justified. Assessing the validity of E-PI, \citet[][p.~3049]{bryan09b} chose not to evaluate the underlying {\it assumptions} in the derivations that yield Eq.~(\ref{tau}). Therefore, even when this equation, or its modifications, are reported to match numerical simulations \citep[it is not always the case, see Fig.~6 of][]{bryan09b}, the nature and  generality of the agreement remain unclear. \citet{zhou2017} in a modelling study showed that with a pronounced sea surface cooling in the eyewall the E-PI assumptions underlying Eq.~(\ref{tau}) (Table~\ref{tabEPI}, second column) do not hold and the cyclone intensity is to a large degree insensitive to the magnitude of the air-sea disequilibrium. Likewise, in Hurricane Isabel 2003 we observe that its intensity ($v_m^2$ varies by $17$\% during the three days of observations) is largely  insensitive to the corresponding variation in $v_E^2$ (varies by two times) and $\mathcal{BC}$ (varies by two times as well), see Table~\ref{tabisab}. In what we consider to be the most transparent discussion of Eq.~(\ref{tau}) to date,
\citet[][p.~173]{emanuel04} justified Eq.~(\ref{tau}) by assuming that at the point of maximum wind $ds/dt$ and $dM/dt$ relate as $\tau_s$ and $\tau_M$, while requiring that $\partial s/\partial z = 0$. However, since at the point of maximum wind $u = 0$ (see appendix~\ref{bryan}), with $\partial s/\partial z = 0$ this means that $ds/dt = 0$, an obstacle not discussed in E-PI derivations \citep[cf.][]{makarieva20}.

Extending E-PI to surface winds,  \citet{re19} did not justify the transition from volume to surface energy fluxes. \citet{ms20} and \citet{makarieva20} pointed out this omission, i.e., the need to explain  how \citet{re19}{\textquoteright}s Eq.~(15) can be derived from their Eq.~(14). In their responses, \citet{re20} and \citet{er20} did not provide the required derivation.  To obtain surface fluxes from volume fluxes, \citet{er20} equated quantities of different dimensions in their Eq.~(6) (and related). This undermines their respective derivations and conclusions. In particular, as demonstrated by \citet{makarieva20}, without an explicit representation of how surface and volume fluxes relate, it is not possible to address the dissipative heating issue, which, as we discuss in  appendix~\ref{bryan}, appears to be responsible for the unexplained singularity in the superintensity account by \citet{bryan09b}  \citep[see also discussions by][]{dhe10,bister11,kieu15,bejan19,er20}.

A systemic problem for the second block of E-PI (Table~\ref{tabEPI}) is that the freedom to  make assumptions about the ratios of surface-to-volume fluxes, is limited. It is the same problem that causes a mismatch between $v_A$ and $v_E$: in E-PI, one cannot freely specify $\partial T/\partial r$. The E-PI assumptions and resulting formulations should be checked for compatibility with two fundamental relationships\footnote{\label{ft4} Dr. Steve Garner suggested that \citet{er20}{\textquoteright}s Eq.~(6) could be fixed if the dimensionless $C_D$ and $C_k$ are re-defined by dividing them by an arbitrary scale height $h$ \citep[cf.][his Eqs.~(8.19) and (8.20)]{emanuel04}. Such a replacement does not indeed change anything in \citet{er20}{\textquoteright}s Eqs.~(2)-(11), but it explicates the incompatibility between their Eqs.~(1) and (2), thus illustrating our argument about conflicting assumptions. Indeed, if we accept \citet{er20}{\textquoteright}s Eq.~(6) with $C_D$ understood as $C_D/h$,  then at the surface $-\mathbf{F}\cdot \mathbf{V} = V^3 C_D /h$. \citet{er20}{\textquoteright}s Eq.~(1) for the isothermal saturated surface flow takes the form of \citet{makarieva20}{\textquoteright}s Eq.~(15), which,
in the present notations,  gives $[\varepsilon/(1-\varepsilon)] L_v dq^*/dt = V^3C_D /h$ after $-\mathbf{F}\cdot \mathbf{V}$ is replaced with $V^3C_D /h$. Comparing this with \citet{er20}{\textquoteright}s Eq.~(2), we find that  the two equations can be reconciled only if $L_v dq^*/dt = V(k_s^*-k_0)C_k /h$, which makes no sense, since the former is the flux of latent heat and the latter -- of enthalpy. Another manifestation of conflicting assumptions related to dissipative heating is the singularity in Eq.~(\ref{brr}) in the superintensity analysis of \citet{bryan09b}, see appendix~\ref{bryan}.}.  The first one is the Clausius-Clapeyron law, which dictates how $dq$, $dT$, $dp$ and $d\mathcal{H}$ relate both on a streamline (where $dp \ne 0$) and across the air-sea interface (where $dp = 0$) \citep[see][their Eq.~(3)]{ar17}. The second one is  the equality between work and turbulent dissipation at the point of maximum wind \citep[see][their Eq.~(9)]{makarieva20}. Besides, as Eq.~(\ref{u}) demonstrates, there are less obvious non-local constraints on the magnitude of the air-sea disequlibrium  at the radius of maximum wind that result from the storm{\textquoteright}s radial and tangential velocity profiles. While a detailed analysis of this subject is beyond the scope of the present paper, we emphasize that the discussion of the numerical validity of E-PI could be more meaningful if  a comprehensive theoretical justification of Eq.~(\ref{tau}) (and its modifications, including the transition between \citet{re19}{\textquoteright}s Eqs.~(14) and~(15) and \citet{er20}{\textquoteright}s Eqs.~(1) and~(2)) were provided.

In the meantime, we conclude by outlining a strategically different perspective on maximum winds. In our view, due to the gross uncertainties surrounding Eq.~(\ref{tau}) and its underlying assumptions, it is difficult to expect that either E-PI, our alternative formulation or any other {\it local} formulation like that of Lilly{\textquoteright}s model, will stand future scrutiny and validate as an informative estimate of storm intensity.
Storm intensity is an integral property of the entire storm{\textquoteright}s energetics,  whereby the power released over a large area is concentrated in the eyewall to generate maximum wind. It cannot be a local function of the highly variable heat input at the radius of maximum wind.  The perceived success of E-PI -- that it produces a plausible if not 100\% robust upper limit on maximum intensity (despite deriving from assumptions that systematically underestimate and overestimate intensities) -- can be explained by the fact that the quantitative parameters in the final expression for kinetic energy incidentally combine into an integral storm parameter -- the partial pressure of water vapor (appendix~\ref{vapor}). Not only the characteristic rates of hurricane {\it intensification} coincide, in their order of magnitude, with precipitation rate \citep{la04}, not only the {\it steady-state} hurricane wind power is proportional to condensation rate \citep{pla15}, but indeed the partial pressure of water vapor is a characteristic scale for maximum kinetic energies observed in real storms. Its exponential temperature dependence would explain that of observed maximum intensities.

\section*{Acknowledgments}
The authors are grateful to Dr. Steve Garner and two anonymous reviewers for their constructive criticisms and suggestions. Our response to the reviewers can be found in appendix~\ref{respr}. Work of A.M. Makarieva is partially funded by the Federal Ministry of Education and Research (BMBF) and the Free State of Bavaria under the Excellence Strategy of the Federal Government and the L\"ander, as well as by the Technical University of Munich -- Institute for Advanced Study.

\appendix
\setcounter{section}{0}%
\setcounter{equation}{0}%
\renewcommand{\theequation}{A\arabic{equation}}%

\section{Deriving the alternative formulation}
\label{altf}

Moist entropy $s$ per unit mass of dry air is defined as (e.g., Eq.~(2) of \citet[][]{em88},  Eq.~(A4) of  
\citet[][]{pa11})
\begin{equation}\label{s}
s = (c_{pd} + q_t c_l) \ln \frac{T}{T'} - \frac{R}{M_d} \ln \frac{p_d}{p'} + q \frac{L_v}{T} - q\frac{R}{M_v}\ln \mathcal{H}.
\end{equation}
Here,  $L_v = L_{v}(T') + (c_{pv} - c_l)(T-T')$ is the latent heat of vaporization (J~kg$^{-1}$); 
$q \equiv \rho_v/\rho_d \equiv \mathcal{H}q^*$ is the water vapor mixing ratio; $\rho_v$ is water vapor density; $\mathcal{H}$ is relative humidity;  $q^* = \rho_v^*/\rho_d$,  $q_l = \rho_l/\rho_d$, and  $q_t = q + q_l$ are the mixing ratio for saturated water vapor, liquid water, and total water, respectively; $\rho_d$, $\rho_v^*$, and $\rho_l$ are the density of dry air, saturated water vapor and liquid water, respectively;  $c_{pd}$ and $c_{pv}$ are the specific heat capacities of dry air and water vapor at constant pressure; $c_l$ is the specific heat capacity of liquid water; $R = 8.3$~J~mol$^{-1}$~K$^{-1}$ is the universal gas constant; $M_d$ and $M_v$ are the molar masses of dry air and water vapor,
respectively;  $p_d$ is the partial pressure of dry air; $T$ is the temperature; $p'$ and $T'$ are reference air pressure and temperature.

For saturated moist entropy $s^*$ ($q = q^*$, $\mathcal{H} = 1$) we have
\begin{gather} \label{Tds2}
Tds^* = (c_{pd} + q_t c_l)dT  - \frac{R T}{M_d} \frac{dp_d}{p_d} + L_v dq^* +q^* dL_v  - q^* L_v \frac{dT}{T}=\left(c_p - \frac{q^* L_v}{T}\right)dT - \frac{R T}{M_d} \frac{dp_d}{p_d} + L_v dq^*  , 
\end{gather}
where $c_p  \equiv c_{pd} + q^* c_{pv} + q_l c_l$ and $dL_v =(c_{pv} - c_l) dT$.  Equation \eqref{Tds2} additionally assumes $q_t = \mathrm{const}$ (reversible adiabat).

The ideal gas law for the partial pressure $p_v$ of  water vapor is
\begin{equation}\label{igv}
p_v = N_v RT,  \quad N_v = \frac{\rho_v}{M_v} , 
\end{equation}
where  $M_v$ and $\rho_v$ are the molar mass and density of water vapor. Using Eq.~(\ref{igv}) with $p_v = p_v^*$ in the
definition of $q^*$ 
\begin{equation}\label{q}
q^* \equiv \frac{\rho_v^*}{\rho_d} = \frac{M_v}{M_d} \frac{p_v^*}{p_d} \equiv \frac{M_v}{M_d} \gamma_d^*,\quad 
\gamma_d^* \equiv\frac{p_v^*}{p_d},
\end{equation}
and  applying the Clausius-Clapeyron law 
\begin{equation}\label{CC}
\frac{dp_v^*}{p_v^*} = \frac{L}{RT} \frac{dT}{T}, \quad L \equiv L_v M_v , 
\end{equation}
we obtain for the last term in Eq.~(\ref{Tds2})
\begin{equation}\label{dq}
L_vdq^* = L_v\frac{M_v}{M_d}\left(\frac{dp_v^*}{p_d} - \frac{p_v^*}{p_d} \frac{dp_d}{p_d}\right)=L_v\frac{M_v}{M_d}\left(\frac{p_v^*}{p_d}\frac{dp_v^*}{p_v^*} - \frac{p_v^*}{p_d} \frac{dp_d}{p_d}\right)
=L_vq^*\left(\frac{L}{RT}\frac{dT}{T} -\frac{dp_d}{p_d}\right) . 
\end{equation}

Using the Clausius-Clapeyron law (\ref{CC}), the ideal gas law $p_d = N_d RT$, where $N_d = \rho_d /M_d$, and noting that $p= p_v^* + p_d$, we obtain for the last but one term in Eq.~(\ref{Tds2})
\begin{equation}\label{pd}
\frac{RT}{M_d}\frac{dp_d}{p_d}=\frac{RT}{M_d} \left(\frac{dp}{p_d} - \frac{dp_v^*}{p_d}\right)  = \frac{dp}{M_dN_d} - \frac{RT p_v^*}{M_d p_d}\frac{dp_v^*}{p_v^*}
=\frac{dp}{\rho_d} - L_v \frac{M_v p_v^*}{M_d p_d}\frac{dT}{T} = \frac{dp}{\rho_d} - q^*L_v\frac{dT}{T}. 
\end{equation}

Taking into account  Eq.~\eqref{pd}, Eq.~\eqref{Tds2} reads
\begin{gather} \label{Tds2ss}
Tds^* =c_p dT - \alpha_d dp + L_v dq^*  . 
\end{gather}
Putting Eqs.~(\ref{dq})  into Eq.~(\ref{Tds2ss}) yields
\begin{gather}\label{Tds3}
Tds^* =\left(c_p +  \frac{L_v q^*}{T} \frac{L(1 + \gamma_d^*)}{RT}\right)dT - \left(1 + \frac{L \gamma_d^*}{RT} \right) \frac{dp}{\rho_d} = 
-(1 +\zeta) \alpha_d dp \left( 1 - \frac{1}{\Gamma}\frac{dT}{dp} \right).
\end{gather}
Here
\begin{equation}\label{Gm}
\Gamma \equiv \frac{\alpha_d}{c_p} \frac{1 + \zeta}{1 +  \mu\zeta (\xi + \zeta)},\quad 
\xi \equiv \frac{L}{RT}, \quad \zeta \equiv \xi \gamma_d^* \equiv \frac{L}{RT}\frac{p_v^*}{p_d} \equiv \frac{L_v q^*}{\alpha_d p_d}, \quad \mu  \equiv \frac{R}{C_p}= \frac{2}{7} , 
\end{equation}
where $\alpha_d \equiv 1/\rho_d$ is the volume per unit mass of dry air and $C_p \simeq c_p M_d$ is the molar heat capacity of air at constant pressure.

Approximating air molar mass by molar mass $M_d$ of dry air  and $c_p$ by $c_{pd}$, we can conveniently express $\Gamma$ as
\begin{equation}\label{Gamma1}
\Gamma \simeq \frac{T}{p} \frac{\mu(1 + \zeta)}{1+ \mu\zeta (\xi + \zeta)} \simeq \frac{T}{p} \frac{\mu(1 + \xi \gamma_d^*)}{1+ \mu \xi^2 \gamma_d^*}.
\end{equation}

E-PI{\textquoteright}s assumption that entropy is well mixed in the boundary layer ($\partial s^*/\partial z = 0$) (block E-II in Table~\ref{tabEPI}) implies a tight link between radial gradients of temperature at a reference height at the surface (the subscript $0$ for temperature-related variables, see Table~\ref{tabEPI}, and $s$ for density and pressure) and at the top of boundary layer (the subscript $b$). When at the radius of maximum wind the surface air is saturated,  as it was, for example, in Hurricane Earl 2010 \citep[][their Table~1]{smith13}, we have $\partial s^*_0/\partial r = \partial s^*_b/\partial r$ and obtain from Eq.~(\ref{Tds3})
\begin{equation}\label{comp}
(1+\zeta_b) \mathcal{C}_b  \frac{\alpha_d}{T_b}\frac{\partial p}{\partial r}\bigg|_{z = z_b} = (1+\zeta_0) \mathcal{C}_0 \frac{\alpha_d}{T_0}\frac{\partial p}{\partial r}\bigg|_{z = 0},
\end{equation}
where $\mathcal{C}$ is defined in Eq.~(\ref{C}).

In hydrostatic equilibrium $\partial p/\partial z \simeq -p/h_d$, $h_d \equiv RT/(gM_d)  \sim 9$~km,
we have for $z_b \ll h_d$
\begin{equation}\label{ptr}
\frac{\partial p}{\partial r}\bigg|_{z = z_b} = \frac{\partial p}{\partial r}\bigg|_{z = 0} + z_b  \frac{\partial^2 p}{\partial z \partial r} \bigg|_{z = 0}  = 
\frac{\partial p}{\partial r}\bigg|_{z = 0} + z_b  \frac{\partial^2 p}{\partial r \partial z} \bigg|_{z = 0}
 \simeq \left(1 - \frac{z_b}{h_d} \right)\frac{\partial p}{\partial r}\bigg|_{z = 0}.
\end{equation}
In the last approximation we have taken into account that $h_d$ can be assumed constant in the boundary layer $z \le z_b \sim 1$~km, since, in the vertical, the relative change of temperature ($\sim 1\%$) is much less than the change of pressure ($\sim 10\%$).

Using Eq.~(\ref{ptr}) and taking into account  that in hydrostatic equilibrium $p_b = (1 - z_b/h_d) p_s$, we have from Eq.~(\ref{comp})  with $\alpha_d \simeq \alpha$
\begin{equation}\label{Cb}
\mathcal{C}_b = \mathcal{C}_0\frac{1+\zeta_0}{1+\zeta_b} \frac{\rho_b}{\rho_s} \frac{T_b}{T_0} \frac{h_d}{h_d - z_b} =
\mathcal{C}_0\frac{1+\zeta_0}{1+\zeta_b} \frac{p_b}{p_s} \frac{h_d}{h_d - z_b} = \mathcal{C}_0\frac{1+\zeta_0}{1+\zeta_b}
= 1.1 \mathcal{C}_0,
\end{equation}
where
\begin{equation}
\zeta_0 =  \dfrac{L}{RT_0} \frac{p_{v0}^*}{p_s} , \quad \zeta_b=\dfrac{L}{RT_b} \frac{p_{vb}^* h_d}{p_s(h_d - z_b)} .
\end{equation}
For characteristic values observed in Hurricane Isabel 2003, $T_0 = 297$~K, $T_0 - T_b = 4$~K we have $p_{v0}^* = 30$~hPa, $p_{vb}^* = 23$~hPa, such that  with $p_s = 10^3$~hPa and $z_b/h_d \simeq 0.1$ the coefficient at $\mathcal{C}_0$ equals $1.1$. (Note that in this evaluation the difference $T_0 - T_b$ is not arbitrary but should correspond to the assumed moist adiabatic lapse rate for $z \le z_b$.) When the surface air is isothermal, $\mathcal{C}_0 = 1$ and $\mathcal{C}_b = 1.1$. This shows that the air temperature at the top of the boundary layer  does increase towards the storm center (this is due to the fact that the water vapor mixing ratio at the surface increases towards the center). However, this increase, and the corresponding $\mathcal{C}_b$ value, are too small to fix the approximately twofold mismatch between $\varepsilon$ and $1/(1 + \zeta)$, see Eq.~(\ref{con}) and Fig.~\ref{fig1}a. That $\mathcal{C}_b$ is greater than $\mathcal{C}_0$ is consistent with the finding of \citet{smith2007} that the horizontal temperature gradient becomes more negative with increasing altitude. \citet{smith2007}{\textquoteright}s Fig.~1b, where $\partial T/\partial r = 0$ at $z \sim 1$~km, corresponds to $\mathcal{C}_b = 1$ and $\mathcal{C}_0 = 0.9$ (a slight decrease of surface air temperature towards the center at the radius of maximum wind).

For an isothermal process with $q_l = 0$ and variable relative humidity, we have from Eq.~(\ref{s}) and $dq/q = (p/p_d)d\mathcal{H}/\mathcal{H} -dp/p_d$
\begin{equation}\label{iTds}
Tds = -\alpha_d dp + L_v dq = -\alpha_d dp \left(1 + \frac{L_v q}{\alpha_d p_d} - \frac{L_v q^* p}{\alpha_d p_d}\frac{d\mathcal{H}}{dp} \right) = -\alpha_d dp \left[ 1 + \zeta \mathcal{H} \left(1 - \frac{p}{\mathcal{H}}\frac{d\mathcal{H}}{dp}\right) \right].
\end{equation} 
Here we omitted the term $-(RT/M_v) \ln \mathcal{H} dq$, which is at least $\xi= L/RT \sim 18$ times less than $L_vdq$. For $\mathcal{C}_0=1$ (horizontally isothermal air at the sea surface), we obtain from Eq.~(\ref{iTds}) and $\partial s_0/\partial r = \partial s^*_b/\partial r$ by analogy with Eq.~(\ref{Cb})
\begin{equation}\label{iCb}
\mathcal{C}_b (1+\zeta_b) = 1+\zeta_0 \mathcal{H} \left(1 - \frac{p_s}{\mathcal{H}}\frac{\partial \mathcal{H}/\partial r}{\partial p_s/\partial r}\right).
\end{equation}
For an adiabatic process with $T_b < T_0$ we have $q^*_b \le q_0 = \mathcal{H}q^*_0 $. With $ \partial \mathcal{H}/\partial r \ge 0$ (relative humidity at the surface increasing towards the storm center) it follows that $\mathcal{C}_b \ge 1$. For $\partial \mathcal{H}/\partial r = 0$,  the maximum value of $\mathcal{C}_b$ corresponds to saturation $\mathcal{H} = 1$ and is given by Eq.~(\ref{Cb}). (For a dry adiabat that only reaches saturation at $z = z_b$ we would have $\mathcal{C}_b \simeq 1$ as $T_b \simeq T_0$). Therefore, E-PI{\textquoteright}s assumptions that the surface air is isothermal \citep[][p.~589]{em86}, while $\partial \mathcal{H}/\partial r = 0$ at the radius of maximum wind \citep[where $\mathcal{H}$ is assumed to be equal to its undisturbed ambient value,][p.~3971]{em95},  are equivalent to assuming $\mathcal{C}_b \simeq 1$ at the point of maximum wind.  If $\partial \mathcal{H}/\partial r \ne 0$, by varying its value it is possible to satisfy Eq.~\eqref{diff} at the point of maximum wind. Then a check of E-PI{\textquoteright}s validity would be not the value of $\partial T_b/\partial r$, but $\partial \mathcal{H}/\partial r$ at the surface, which, in this context, cannot be freely specified in E-PI. However, for stronger storms that reach saturation at the radius of maximum wind, we would have $\partial \mathcal{H}/\partial r = 0$ (section~\ref{dTdr}).

\setcounter{equation}{0}%
\renewcommand{\theequation}{B\arabic{equation}}%

\section{Hypercanes}
\label{hyper}

Hypercanes were introduced as winds with theoretically infinite velocities that should occur with sea surface temperatures  exceeding approximately 40{\textcelsius} \citep{em88}, i.e., at those temperatures where $1/\varepsilon = 1 + \zeta$ and the solid and dashed lines in Fig.~\ref{fig1}a begin to intersect. It is not a coincidence.

The singularity responsible for hypercanes first appeared in \citet{em86}{\textquoteright}s Eq.~(26) for the central pressure drop. This equation derives from combining two equations.  The first one is Eq.~(\ref{iTds}) for the horizontally isothermal air at the sea surface, which corresponds to \citet{em86}{\textquoteright}s Eq.~(25). The second one is $\varepsilon T_b ds^* = - \alpha_d dp$ for the top of the boundary layer $z = z_b$, which corresponds to \citet{em86}{\textquoteright}s Eq.~(21). From these two equations, assuming as before that  $\partial s^*_0/\partial r = \partial s^*_b/\partial r$ and using Eq.~(\ref{ptr}) \citep[cf. the unnumbered equation after Eq.~(25) on p.~589 of][]{em86}, we obtain 
\begin{equation}\label{eq26}
\frac{1}{\varepsilon} =  1+\zeta_0 \mathcal{H} \left(1 - \frac{p_s}{\mathcal{H}}\frac{\partial \mathcal{H}/\partial r}{\partial p_s/\partial r}\right).
\end{equation}
Solving this for $(1/p_s)\partial p_s/\partial r$ gives
\begin{equation}\label{eq26!}
\frac{\partial \ln p_s}{\partial r } = -\frac{\varepsilon \zeta_0 \partial \mathcal{H}/\partial r}{1 - \varepsilon (1 + \zeta_0 \mathcal{H})}.
\end{equation}
Linearizing this yields \citet{em86}{\textquoteright}s Eq.~(26) (where in the numerator the last term proportional to squared outflow radius is for simplicity omitted).

The singularity corresponding to $\partial p_s/\partial r \to \infty$ arises at $1/\varepsilon = 1 + \zeta_0 \mathcal{H}$
under the assumption that $\varepsilon$ is independent of $\zeta_0$. This assumption is incorrect, since for E-PI to be valid, Eq.~(\ref{con}) must hold, such that $1/\varepsilon = \mathcal{C}_b (1 + \zeta_b)$.  In view of Eq.~(\ref{iCb}), the latter expression is equal to the right-hand part of Eq.~(\ref{eq26}), which means that \citet{em86}{\textquoteright}s Eq.~(26) is an identity, from which nothing can be deduced. Hypercanes do not exist.

\setcounter{equation}{0}%
\renewcommand{\theequation}{C\arabic{equation}}%

\section{The superintensity analysis of \citet{bryan09b}}
\label{bryan}

\citet{bryan09b} intended to derive a theoretical expression for maximum potential intensity that would be valid under the same assumptions as E-PI except for the gradient-wind and hydrostatic balances. The result was their Eq. (24):
\begin{equation}\label{br}
v_{B}^2 = v_E^{*2} + a r_m \eta_m w_m.
\end{equation}
Here $v_E^*$ is a E-PI velocity estimate that involves dissipative heating ($v_E^{*2} = a v_E^2$ with $a \equiv T_s/T_o > 1$),
$w$ is vertical velocity, $\eta \equiv \partial u/\partial z - \partial w/\partial r$  is the azimuthal component of absolute vorticity, and the subscript $m$ indicates that the variables are evaluated at the point of maximum tangential wind.

To derive Eq.~(\ref{br}), \citet[][their Eqs.~B1 and B2]{bryan09b} considered the equations of motion for radial and vertical velocities, but omitted to consider the tangential velocity equation. From this equation for an axisymmetric inviscid flow \citep[see, e.g., Eq.~(A10b) of][]{mpi4-jas}
\begin{equation}\label{eqv}
u\left[f + \frac{1}{r}\frac{\partial (r v)}{\partial r}\right] + w \frac{\partial v}{\partial z}=0,
\end{equation}
it follows that, at the point of maximum tangential wind, the radial velocity is zero. This condition may be {\it approximate} in a numerical model that {\it violates E-PI{\textquoteright}s assumption of zero viscosity at the point of maximum wind}. But it must be {\it exactly} obeyed by any theoretical derivation that, as in the analysis of
\citet{bryan09b}, keeps this assumption intact.

With $u=0$, the equation of motion for radial velocity, \citet{bryan09b}{\textquoteright}s Eq.~(21), becomes their Eq.~(22)
\begin{equation}\label{br2}
\eta_m w_m = \frac{v_m^2}{r_m} - \alpha \frac{\partial p}{\partial r} \equiv (1 - \mathcal{B})\frac{v_m^2}{r_m}
\end{equation}
under an additional {\it minor} assumption
that $|\eta| \equiv |\partial u/\partial z - \partial w/\partial r| \gg |\partial w/\partial r|$. The identity uses our definition of $\mathcal{B}$ (\ref{gwbg}).

From the law of conservation of mass, the latter  condition should be fulfilled when $(z_m/\Delta r_m)^2 \ll 1$, where $z_m$
is the altitude of the point of maximum tangential wind (where $u = 0$) and $\Delta r_m$ is the radial half-width of the eyewall.
Under realistic conditions, this relationship should always hold. For example, for \citet{bryan09b}{\textquoteright}s control simulation with $z_m \simeq 1$~km and $\Delta r_m \simeq 7$~km (retrieved from their Fig.~3),  the inaccuracy of Eq.~(\ref{br2}) due to the neglect of $\partial w/\partial r$ as compared to $\eta$ should be of the order of 2\%. This appoximately agrees with the available information about the control simulation: $\eta_m = 0.03$~s$^{-1}$ \citep[][p.~3055]{bryan09b} and $|\partial w/\partial r| < \Delta w/\Delta r_m = 0.001$~s$^{-1}$, where we calculated $\Delta w = w_m - w = 7.5$~m~s$^{-1}$ taking into account that $w=0.5$~m~s$^{-1}$ at the eyewall outer borders and $w_m=8$~m~s$^{-1}$ at the point of maximum wind \citep[see, respectively,][their Fig.~3 and p.~3055]{bryan09b}.

From Eq.~(\ref{br2}) we have
\begin{equation}\label{B}
\mathcal{B} = 1 - \frac{r_m \eta_m w_m}{v_m^2}.
\end{equation}
Based on the data of \citet{bell08} for Hurricane Isabel 2003, \citet{bryan09b} compiled terms from the right-hand part of Eq.~(\ref{B}) in their Table~1. Using that table, we estimate $\mathcal{B}$ for September 12, 13 and 14 as, respectively, $0.95$, $0.72$ and $0.85$  (these values are used in our Table~\ref{tabisab}). For September 13, $\mathcal{B}=0.72$ obtained  from Eq.~(\ref{B}) approximately agrees with the statement of \citet[][p.~2037]{bell08} that on this day {\textquotedblleft}the boundary layer tangential wind was $\sim15\%$ supergradient{\textquotedblright} near the radius of maximum wind (this corresponds to $\mathcal{B} = 1/1.15^2 = 0.76$).
For their control simulation, \citet[][p.~3055]{bryan09b} report $v_m = 109$~m~s$^{-1}$, $r_m = 17.5$~km, $\eta_m = 0.03$~s$^{-1}$, $w_m = 8$~m~s$^{-1}$. Using these values, we estimate $\mathcal{B} = \mathcal{B}_{\rm 1} = 0.65$ from Eq.~(\ref{B}).

Using Eq.~(\ref{B}) in Eq.~(\ref{br}), while assuming that $v_{B} = v_m$, we find
\begin{equation}\label{brr}
v_{B}^2 = \frac{v_E^{*2}}{1 - a (1 - \mathcal{B})}.
\end{equation}
Putting $\mathcal{B}= \mathcal{B}_{\rm 1}=0.65$ into Eq.~(\ref{brr}) with $a = 1.5$  and $v_E^* = 72$~m~s$^{-1}$ (the values \citet{bryan09b} report for their control simulation), we find $v_{B1} = 105$~m~s$^{-1}$. This is in good agreement with $v_{B} = 107$~m~s$^{-1}$ obtained by \citet{bryan09b} from their original formulation given by Eq.~(\ref{br}). The minor numerical discrepancy is due to $v_{B} = v_m$ that we put to obtain Eq.~(\ref{brr}) from Eq.~(\ref{B}). (It was justified, since $v_{B}$ (\ref{br}) is a theoretical expression for $v_m$ that is expected to be valid in an inviscid atmosphere where Eq.~(\ref{B}) also holds.)

From the fact that their estimated $v_{B}= 107$~m~s$^{-1}$ is close to the actual maximum velocity $v_m = 109$~m~s$^{-1}$ in their control simulation, \citet[][p.~3055]{bryan09b} concluded that {\textquotedblleft}the neglect of unbalanced flow effects is mostly responsible for the systematic underprediction by E-PI{\textquotedblright}. However, such a close match between the theoretical expression  and control simulation looks unexpected. How could \citet{bryan09b}{\textquoteright}s derivation have produced an estimate of $v_m$ accurate to within a couple of percent, if one of E-PI{\textquoteright}s key equations, on which this derivation is based, errs  in their control simulation by  a factor of $1.5$ (see \citet{bryan09b}{\textquoteright}s Eq.~8, Fig.~6 and p.~3049, and discussion below)? It could only be if one major discrepancy compensated another.

\begin{figure*}[tbp]
\begin{minipage}[p]{0.85\textwidth}
\centering\includegraphics[width=0.7\textwidth,angle=0,clip]{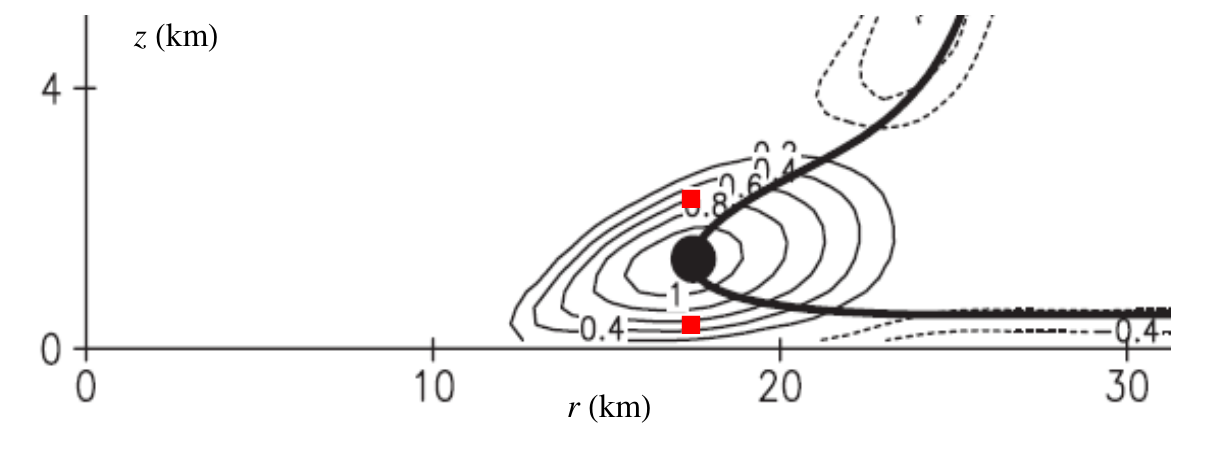}
\end{minipage}
\caption{Part of \citet{bryan09b}{\textquoteright}s Fig.~8 with the  thick curve showing the streamline to which the point of maximum wind (black circle) belongs, and the thin contours showing $(1-\mathcal{B})/\mathcal{B}$ with the interval of $0.2$, in \citet{bryan09b}{\textquoteright}s control simulation. We have added two red squares to indicate possible locations where the point of maximum wind should have been if $\mathcal{B} = \mathcal{B}_{\rm 1} = 0.65$  and $(1-\mathcal{B})/\mathcal{B} = 0.55$ corresponding to the data from \citet{bryan09b}{\textquoteright}s section~6b, see Eq.~(\ref{B}), were correct.}
\label{figbryan}
\end{figure*}

Indeed, a closer inspection of these results reveals that they are not self-consistent.
In their Fig.~8, \citet{bryan09b} analyzed the degree to which the gradient-wind balance is broken in their control simulation by plotting contours of $(1 - \mathcal{B})/\mathcal{B}$ (Fig.~\ref{figbryan}). That figure shows that the point of maximum wind at $r_m = 17.5$~km corresponds to $(1 - \mathcal{B})/\mathcal{B} \gtrsim 1$ and $\mathcal{B} = \mathcal{B}_{\rm 2} \lesssim 0.5$. This is confirmed in the text, which says
that at the point of maximum wind the sum $v_m^2/r_m + fv_m$ (here the second term is negligibly small)
is twice the absolute magnitude of $\alpha \partial p/\partial r$ \citep[see][p.~3050 and their Eq.~(12)]{bryan09b}.

Given that Eq.~(\ref{brr}) has a singularity at $\mathcal{B} = 1-1/a \simeq \varepsilon$, the estimate of $v_{B}$ is sensitive to minor variations around $\mathcal{B} = 1-1/a = 0.33$, to which $\mathcal{B}_{\rm 2} \lesssim 0.5$, that is retrieved from Fig. 8 of \citet{bryan09b}, is sufficiently close. Putting $\mathcal{B} = \mathcal{B}_{\rm 2} = 0.5$ into Eq.~(\ref{brr}) we obtain $v_{B2} = 144$~m~s$^{-1}$, which is far beyond  $v_m = 109$~m~s$^{-1}$ in their control simulation. Until this major discrepancy, $\mathcal{B}_{1} \ne \mathcal{B}_{2}$ and $v_{B1} \ll v_{B2}$, is resolved (Fig.~\ref{figbryan}), \citet{bryan09b}{\textquoteright}s conclusion, that their numerical simulations support the hypothesis of superintensity being largely due to supergradient winds, lacks a solid ground.

Furthermore, the infinite winds at $\mathcal{B} = 1-1/a > 0$, implied by their derivations and exposed by Eq.~(\ref{brr}), require an explanation.
While we ultimately leave it to the authors to discuss this  yet another E-PI singularity (the first one being $1/[1 - \varepsilon(1+\zeta)]$ in \citet{em86}{\textquoteright}s Eq.~(26), see section~\ref{isab} and appendix~\ref{hyper}), here we can offer one thought. That $v_{B} \to \infty$ at a finite $\mathcal{B} \to 1-1/a$ is solely due to  $a > 1$. The latter inequality is a consequence of {\textquotedblleft}dissipative heating{\textquotedblright} assumed to recirculate within the cyclone to make it stronger. Without this dissipative heating, $a = 1$ in Eq.~(\ref{brr}) \citep[see footnote 2 of][]{bryan09b}, and the unphysical singularity disappears. The {\textquotedblleft}superintensity{\textquotedblright} effect of supergradient winds is then accounted for by dividing the conventional ({\textquotedblleft}balanced{\textquotedblright}) $v_E^2$ (\ref{vE}) by $\mathcal{B}<1$, a straightforward procedure identical to what we applied in our alternative formulation in Eq.~(\ref{sr1}), see also Eq.~(37) of \citet{mpi4-jas}. 

Along with their control simulation, \citet[][their Fig.~12]{bryan09b} reported additional results intended to illustrate a good agreement between $v_B$ (\ref{br}) and $v_m$ across a range of horizontal mixing lengths $l_h$. However, they did not indicate whether in these additional simulations the atmosphere is inviscid at the point of maximum wind, as it approximately is in their control simulation \citep[][p.~3050]{bryan09b}. Until this is demonstrated (which will however mean that for these simulations Eq.~(\ref{br2}) also holds with a high precision), the data of  \citet{bryan09b} cannot be interpreted as supporting their Eq.~(\ref{br}), which was derived for an inviscid atmosphere. It should also be noted that while \citet[][]{bryan09b}{\textquoteright}s Fig.~12 indicates a very close (to within a couple of percent) agreement between $v_B$  (\ref{br}) and $v_m$  for $47~{\rm m} \le l_h \le 1500~{\rm m}$, subsequent analysis did not confirm such a close match but revealed that for $l_h \le 300$~m there is a 20\% superintensity unexplained by $v_B$ \citep[see][p.~1137 and their Fig.~13 for $C_k/C_d = 1$ and $l_v = 200$~m]{bryan12}.

There is another reason why the claim for generality in \citet{bryan09b}{\textquoteright}s Fig.~12 warrants caution. \citet{bryan09a}{\textquoteright}s simulations used in \citet{bryan09b}{\textquoteright}s Fig.~12 use a different {\it vertical} mixing
length $l_v = 200$~m as compared to $l_v = 100$~m in \citet{bryan09b}{\textquoteright}s  control simulation. According to \citet[][see their Fig.~2 and p.~1777]{bryan09a}, varying  $l_v$ makes virtually no impact on maximum velocity over a range from $l_v = 25$~m to $l_v = 400$~m. At the same time though, this parameter is apparently instrumental in bringing E-PI{\textquoteright}s assumptions in agreement with the numerical model.  Indeed, one of E-PI{\textquoteright}s key assumptions, \citet[][]{bryan09b}{\textquoteright}s Eq.~(8) ({\textquotedblleft}the fundamental closure{\textquotedblright}),  is violated by 50\% in the control simulation with $l_v = 100$~m, but with $l_v = 200$~m it becomes almost perfectly accurate  \citep[cf. Figs.~6 and 7 of][]{bryan09b}. \citet[][]{bryan09b} tentatively attributed this to the boundary layer being {\textquotedblleft}better resolved{\textquotedblright} with $l_v = 200$~m. If correct, for this to be accepted as a general explanation, an analysis over a larger range of $l_v$ would seem to be required. Whether the validity of E-PI{\textquoteright}s dissipative heating formulation hinges on a model parameter that does not matter for maximum velocity, and if yes, what the implications are, remains to be investigated.

We conclude by demonstrating how our alternative expression (\ref{sr1}) can be applied to the control simulation of  \citet{bryan09b}. Dissipative heating in E-PI is effectively accounted for by multiplying $\tau_s/\tau_M$ in  Eq.~(\ref{tau}) by  the factor of  $a \equiv T_s/T_o > 1$, to obtain $v_E^{*2} = av_E^2$ instead of $v_E^2$ in Eq.~(\ref{vE}).  However, E-PI{\textquoteright}s dissipative heating equation, i.e., \citet{bryan09b}{\textquoteright}s Eq.~(8), is strongly violated in their control simulation.  The ratio between the left-hand and right-hand sides of this equation, instead of being unity, is precisely equal to $1/a = 1/1.5$ \citep[see][p.~3049 and their Fig.~6]{bryan09b}.  This means that \citet{bryan09b}{\textquoteright}s Eq.~(8) without dissipative heating -- i.e., our relationship (\ref{tau}) -- for their control simulation is exact. In consequence, E-PI{\textquoteright}s maximum velocity diagnosed from our Eq.~(\ref{dsE2}) in \citet{bryan09b}{\textquoteright}s control simulation should be $v_E^*/\sqrt{a}$. It would be $v_E^*$ if E-PI{\textquoteright}s dissipative heating equation -- \citet{bryan09b}{\textquoteright}s Eq.~(8) -- were accurate.\footnote{It is noteworthy that \citet{bryan09b} chose not to investigate the numerical consequences of the gross inaccuracy of their Eq.~(8) on the grounds that it overestimates the actual $v_E$  and thus cannot explain superintensity (i.e., it cannot explain the E-PI estimate being too low). But this overestimate implies that the degree of superintensity to be explained is larger than what follows from E-PI{\textquoteright}s dissipative heating equation (\citet{bryan09b}{\textquoteright}s Eq.~(8))  used in their formula for $v_B$ (\ref{br}). This is what Eq.~(\ref{brr}) with $\mathcal{B}=\mathcal{B}_2$ essenitally exposed.}

With $a (v_m/v_E^*)^2 = 1.5 (109/72)^2 = 3.4$, we have a significant superintensity.
With $\mathcal{B} = 0.5$, from Eqs.~(\ref{sr1}) and (\ref{cf}) we find that $[(1+\zeta)\varepsilon \mathcal{C}]^{-1} = \mathcal{B} (v_m/v_E)^2 = 1.7$. This means that supergradient winds account for one half of the superintensity factor $3.4$,  leaving the remaining $1.7$ unexplained. For $\varepsilon \simeq 1 - 1/a = 0.33$ assuming $T_b = 293$~K as in Hurricane Isabel 2003, which had the temperature of surface air (297.5~K) similar to \citet{bryan09b}{\textquoteright}s control simulation (298~K), we have  $1/(1+\zeta) = 0.67$ for $850$~hPa and $\mathcal{C} = [\mathcal{B} (v_m/v_E)^2 (1+\zeta)\varepsilon]^{-1} = 1.2$. This situation is comparable to Hurricane Isabel 2003 on September 12 (Table~\ref{tabisab}), i.e.,  the air at the point of maximum wind in \citet{bryan09b}{\textquoteright}s control simulation is close to horizontal isothermy. This explains why in \citet{bryan09b}{\textquoteright}s control simulation E-PI strongly underestimates $v_m$ even after the account of supergradient winds.  \citet{wang2014}, who followed \citet{bryan09b}{\textquoteright}s approach, similarly found that the effect of supergradient winds was insufficient to explain superintensity in their 3D model. With no other quantitative explanation at hand, they hypothesized that the discrepancy could be attributed to the neglect of turbulent mixing or to cyclones not being truly in steady state, but did not examine $\partial T_b/\partial r$.

\setcounter{equation}{0}%
\renewcommand{\theequation}{D\arabic{equation}}%

\section{E-PI and the partial pressure of water vapor}
\label{vapor}

Using Eq.~(\ref{q}) and $q \equiv \mathcal{H} q^*$, where $\mathcal{H} \equiv p_v/p_v^*$ is relative humidity, 
the E-PI upper limit $\hat{v}_E$ on maximum velocities (Table~\ref{tabEPI}) can be re-written as follows:
\noindent
\begin{equation}\label{sc1}
\hat{v}_E^2 \sim \left(\frac{\varepsilon}{2}\frac{C_k}{C_D}\frac{L}{RT_s} \frac{1 - \mathcal{H}_a}{\mathcal{H}_a}\right)
\frac{2 p_{va}}{\rho_{a}} .
\end{equation}
Here $p_{va} = \mathcal{H}_ap_{va}^*$ is the actual partial pressure of water vapor in surface air in the ambient environment, $p_{va}^*$ is the  partial pressure of saturated water vapor at sea surface temperature in 
the ambient environment, $\rho_{a}$ is ambient air density at the surface. Using typical tropical values $T_s = 300$~K, $\mathcal{H}_a = 0.8$, $\rho_{a} \simeq 1.2$~kg~m$^{-3}$,  $\varepsilon = 0.32$ \citep[Table~1 of][]{em89} 
and $C_k/C_D = 1$, we have $p_{va} = 28$~hPa and $v_{m} \simeq 60$~m~s$^{-1}$ in agreement with Table~1 of \citet{em89}.

The coefficient in parentheses in Eq.~(\ref{sc1}) for the same typical parameters is close to unity and depends only weakly on air temperature:
\begin{equation}\label{fact}
\frac{\varepsilon}{2} \frac{C_k}{C_D} \frac{L}{RT_s} \frac{1 - \mathcal{H}_a}{\mathcal{H}_a} \simeq 1.
\end{equation}
This means that numerically the scaling of maximum velocity in E-PI practically coincides with the scaling
\begin{equation}\label{osc}
\rho_{a} \frac{\hat{v}_E^2}{2} = p_{va}
\end{equation}
proposed within the concept of condensation-induced atmospheric dynamics \citep[for a more detailed discussion see][section~5]{makarieva18b}.  Introducing the dissipative heating leads to additional factor $1/(1-\varepsilon) \sim 1$ in Eq.~(\ref{fact}) and $\hat{v}_E^2$.

\section{Response to the reviewers}
\label{respr}

{\color{blue}\it
\noindent
Dec 01, 2021

\noindent
Ref.: JAS-D-21-0149

\noindent
Editor Decision

\vspace{0.45cm}\noindent
Dear Dr. Makarieva,

\vspace{0.15cm}
I am now in receipt of all reviews of your manuscript {\textquotedblleft}Alternative expression for the maximum potential intensity of tropical cyclones{\textquotedblright}, and an editorial decision of Major Revision has been reached. The reviews are included below or attached.

Although Rev 1 recommended accept, the reviewer also pointed out that the study is {\textquotedblleft}incremental{\textquotedblright} instead of  {\textquotedblleft}revolutionary{\textquotedblright}, and that the significance of the study is likely overstated due to the lack of a proper context. Additionally, Rev 2 questioned the validity of some arguments. Since the manuscript is a resubmission and has gone through two rounds of review, I hope that the reviewers' comments will be sufficiently addressed during revision so that the manuscript can be moved forward.

We invite you to submit a revised paper by Jan 30, 2022. If you anticipate problems meeting this deadline, please contact me as soon as possible at ZWang.JAS@ ametsoc.org.

Along with your revision, please upload a point-by-point response that satisfactorily addresses the concerns and suggestions of each reviewer. To help the reviewers and Editor assess your revisions, our journal recommends that you cut-and-paste the reviewer and Editor comments into a new document. As you would conduct a dialog with someone else, insert your responses in a different font, different font style, or different color after each comment. If you have made a change to the manuscript, please indicate where in the manuscript the change has been made.  (Indicating the line number where the change has been made would be one way, but is not the only way.)

Although our journal does not require it, you may wish to include a tracked-changes version of your manuscript. You will be able to upload this as {\textquotedblleft}additional material for reviewer reference{\textquotedblright}. Should you disagree with any of the proposed revisions, you will have the opportunity to explain your rationale in your response.

Please go to www.ametsoc.org/PUBSrevisions and read the AMS Guidelines for Revisions. Be sure to meet all recommendations when revising your paper to ensure the quickest processing time possible.

When you are ready to submit your revision, go to https://www.editorialmana\-ger.com/amsjas/ and log in as an Author. Click on the menu item labeled {\textquotedblleft}Submissions Needing Revision{\textquotedblright} and follow the directions for submitting the file.

Thank you for submitting your manuscript to the Journal of the Atmospheric Sciences. I look forward to receiving your revision.

\vspace{0.45cm}\noindent
Sincerely,

\vspace{0.35cm}
\noindent
Dr. Zhuo Wang 

\noindent
Editor

\noindent
Journal of the Atmospheric Sciences
}

\newpage
\noindent
January 25, 2022

\noindent
Ref.: JAS-D-21-0149

\noindent
Resubmission of revised manuscript\footnote{Manuscript JAS-D-21-0149R with line numbers can be found at \href{https://bioticregulation.ru/ab.php?id=alt}{https://bioticregulation.ru/ab.php?id=alt}}

\vspace{0.15cm}\noindent
Dear Editors,

\vspace{0.15cm}\noindent
Thank you for your consideration of our work {\it {\textquotedblleft}Alternative expression for the maximum potential intensity of tropical cyclones{\textquotedblright}}, 
which we are now re-submitting having carefully addressed our reviewers' additional comments.

\vspace{0.15cm}\noindent
We sincerely appreciate that, after rigorous examinations, two experts have found sufficient merit in our work to recommend it for publication, 
and that even our most critical reviewer, rather than criticizing our main result, now mostly focus their efforts on defending the study of \citet{bryan09b}.

\vspace{0.15cm}\noindent
In our work we present a new formulation of maximum potential intensity that reveals how E-PI superintensity can result from the horizontal isothermy at the point of maximum wind. This is a new finding that, as admitted by all the reviewers, pertains to an important and complex topic.  Additionally, our work discusses several previously unrecognized essential incorrectnesses:  an incoherence in \citet{er20}{\textquoteright}s assumptions related to dissipative heating (JAS 77, 3977 (2020)), hypercanes (JAS 45, 1143 (1988)), and  an incoherence in the theoretical and numerical analyses in the superintensity study of \citet{bryan09b} (JAS 66, 3042 (2009)).

\vspace{0.15cm}\noindent
The latter two issues were raised via the mediation of our reviewers, who have significantly contributed to shaping the present manuscript. We
gratefully acknowledge these inputs. At the current stage, however, we respectfully disagree with our second reviewer about the need  for a further major revision of our work on the part of our assessment of \citet{bryan09b}{\textquoteright}s analyses. The reviewer purported to defend \citet{bryan09b}{\textquoteright}s results by noting that their conclusions are more general than their control simulation and that they do not involve their Eq. (22). As we show in our response, these statements are incorrect. No major revision of any of our arguments was therefore necessary. (We show that the discrepancy in the results of \citet{bryan09b} could have remained unnoticed due to an incomplete analysis of the equations of motion and due to neglected consequences of E-PI{\textquoteright}s dissipative heating formulation being strongly violated in their control simulation.)

\vspace{0.15cm}\noindent
This said, we have made every effort to ensure, taking into account our reviewer{\textquoteright}s detailed comments, that our characterization of \citet{bryan09b}{\textquoteright}s work is as accurate and clear as possible. This involved a few additional clarifications and minor modifications in appendix~C. We hope that, in this thoroughly verified form, our work could be suitable for publication.  As urged by the AMS publishing policy, we believe that the revealed issues require a thorough evaluation by the broader community.

\vspace{0.15cm}\noindent
Sincerely,

\vspace{0.15cm}\noindent
Anastassia Makarieva

\vspace{0.15cm}\noindent
Andrei Nefiodov

\vspace{0.5cm}\noindent
RESPONSE TO THE REVIEWERS

\vspace{0.15cm}\noindent
We sincerely thank our reviewers for their additional time and efforts.

\vspace{0.5cm}\noindent
{\color{blue} \it
\noindent
REVIEWER COMMENTS
\vspace{0.15cm}

\vspace{0.15cm}
\begin{center}
Reviewer \#1: 
\end{center}

\vspace{0.15cm}\noindent
I admit to having some review fatigue at this point, and I don't find any math errors in the manuscript, so permit me to stick to the larger issues.  First of all, the manuscript is a deep dive into an issue that I am quite sure many analytical experts have considered but never pursued.  I credit the authors with doing the hard work of pursuing the implications meticulously and raising awareness, but I still don't consider the investigation to be revolutionary, as opposed to just good incremental science.  This has been my overarching problem with the writing, starting with the title.  This is NOT an alternative MPI because it is not well enough constrained and in fact I do NOT believe there can exist a one-dimensional model of a TC, which would mean that the rest of the atmosphere is irrelevant.  The authors state that V and (presumably) p maps are not very useful.  I find this astonishingly brazen if it is based entirely on their 1-D analysis.  Of course there
are problems with the full energy-cycle analysis because of difficulties coming up with constraints and approximations, but one cannot dismiss the voluminous literature investigating these problems in the 2D context.  The authors have cut the energy cycle into 2 pieces, which is the reason why dk (change in moist enthalpy) re-appears in the integral, focusing attention on dT/dp as a flag for unphysical aspects of the traditional analysis.  This is the main contribution of the paper and a good one.  The authors seem to resist this simple way to explain what they are doing with their 1-D model.  Why?   (Note that -dk appears in the other half of the energy cycle.)  I understand that lack of context like this makes the paper seem more revolutionary, but is it worth it?  It looks to me like C~$\sim 2$ recovers the traditional 0.3 Carnot efficiency pretty closely.  But the efficiency depends on the effective cooling temperature in the energy cycle and it has always been suspected that
this effective temperature is too low and the efficiency is too large because of updraft entrainment.  Recently, the literature prefers the term {\textquotedblleft}ventilation{\textquotedblright} (Tang and Emanuel).  I don't want to be too specific about all the problems with the energy cycle.   I just want to point out that the authors' analysis is just one of the ways to focus attention on these problems.  It might even be the best way.   I think the paper at this point will rise or fall on the tone the authors adopt, but I'm old-school about how to advance science.  One doesn't try to re-invent the wheel and and make it sound like everything that others have done is {\textquotedblleft}not very useful.{\textquotedblright}  Good luck.

\vspace{0.15cm}\noindent 
{\color{black}\rm	
Thank you for sharing your further insights. We did not use the phrase {\it "not very useful"}, and we would have
never used it to refer to other people{\textquoteright}s efforts. What we said in the manuscript, exactly, was: {\it "First, calculating $\hat{v}_E$ (Table~\ref{tabEPI})  for an arbitrary environment may not be very informative: if the cyclone modifies its outflow region,
there should exist {\textquotedblleft}unmodified{\textquotedblright} environments where cyclonic outflows may never happen."}  To interpret this as {\color{blue}\it "the authors state that V and (presumably) p maps are not very useful"} appears to be a misunderstanding.

\vspace{0.15cm}\noindent 
We likewise do not feel that we can be fairly accused of trying to make our work look {\color{blue}\it "more revolutionary"}, especially as we note in our conclusions that, in our opinion, none of the existing local formulations of maximum potential intensity (including ours) {\it "will stand future scrutiny and validate as an informative estimate of storm intensity"}.  Rather, we agree with the reviewer that the merit of our work is probably indeed in raising {\color{blue}\it "a flag for unphysical aspects of the traditional analysis"}.

\vspace{0.15cm}\noindent 
In this well-defined context, which already the first sentence of our abstract introduces the reader to, we believe that the title of our work is accurate. We agree that our formulation {\color{blue}\it "is not well enough constrained"}, but it is  an {\it alternative} to E-PI, which, as our work clarifies, is not well enough constrained either. Our formulation has a {\it not well enough constrained} parameter $\mathcal{C}$ that reflects the horizontal temperature gradient in the boundary layer\footnote{To be precise, one more parameter that is not explicitly constrained is the air pressure $p_d$ at the point of maximum wind that enters the definition  of $\gamma_d^*$. But, as Fig.~\ref{fig1}a shows, for realistic values of $p_d$ its influence on $v_A$ is minimal.  As a compensation for this minor additional uncertainty, $v_A$ is more robust than $v_E$, i.e., it is based on fewer assumptions.}. E-PI, in its turn, has a {\it not well enough constrained} parameter $T_o$, the temperature where $v = 0$ for the adiabatic outflow.  Inasmuch as the reviewer believes that our formulation implies {\color{blue}\it "that the rest of the atmosphere is irrelevant"}, one can say that E-PI ignores all processes in the boundary layer that may impact $\partial T/\partial r$ (i.e., $\mathcal{C}$) at the point of maximum wind, as if they were controlled by the outflow temperature. These neglected processes include, importantly, the temperature gradients generated by air motion at the surface, as described by our Eq.~(\ref{u}).

\vspace{0.15cm}\noindent 

Given that $\mathcal{C} = 1$ corresponds to a twofold underestimate, it is true indeed that {\color{blue}\it "C~$\sim 2$ recovers the traditional 0.3 Carnot efficiency pretty closely"}. But why should $\mathcal{C}$ be around two? Regarding that {\color{blue}\it "it has always been suspected that this effective temperature is too low and the efficiency is too large because of updraft entrainment"}, as we note in the revised text, the ventilation could explain why an observed or modeled intensity is {\it lower} than E-PI. We are concerned about explaining {\it superintensity} -- the opposite effect corresponding to $\mathcal{C} = 1$. We show how it can be explained by stratospheric warming violating adiabaticity of the outflow using the framework of \citet{mpi4-jas}. To our knowledge, this has not been pointed out before.

\vspace{0.15cm}\noindent 
While we do regret that the reviewer appears to remain unsatisfied with some of the more subjective aspects of our work like, e.g., its tone,
we highly appreciate the reviewer{\textquoteright}s strict objectivity in judging its scientific content. Thank you again.
}


\vspace{1cm}
\begin{center}
Reviewer \#2: 
\end{center}

\vspace{0.15cm}
\noindent
\textbf{Recommendation:} Major Revisions

\vspace{0.15cm}
\noindent
\textbf{Summary:} 
This manuscript has again been improved from the previous version. Aside from my continued disagreement with the authors' primary conclusions (as described in detail in my previous reviews), my most important major concerns with the previous version of this manuscript were that it still made several arguments that misleadingly implied that E-PI was reliant on assumption it doesn't actually make, and that the authors had not adequately dealt with previous studies that had shown superintensity can be largely explained as a consequence of unbalanced flow. Although there are still some characterizations of E-PI that I would quibble with, for the most part, I think the authors have addressed the former concern, and I am also basically ok with their characterization of my point of view that they have now incorporated into their revised introduction. With respect to the latter concern, I appreciate that the authors have now attempted to reckon with the key study of BR09 and its finding that when diagnostically accounting for unbalanced flow, this can already explain superintensity. However, I think the authors' analysis and critique of BR09 is flawed (for reasons detailed below), and so I still don't think the manuscript is ready to be published. Therefore, I am again recommending Major Revisions.

\vspace{0.15cm}\noindent 
{\color{black}\rm  We are grateful to the reviewer for all the detailed discussions, which we have greatly appreciated.
However, as we explain below, the reviewer{\textquoteright}s criticisms of our assessment of
\citet{bryan09b}{\textquoteright}s study are ungrounded.}

\vspace{0.15cm}
\noindent
\textbf{Major Comment:}
A. Is there a major discrepancy in the analysis of BR09?

\vspace{0.15cm}\noindent 
{\color{black}\rm
Yes, there is. That the reviewer has been able to conclude otherwise is, in our opinion, related to the incomplete analysis of the equations of motion by \citet{bryan09b},
as demonstrated below.} 

\vspace{0.15cm}
\noindent
The authors attempt to demonstrate (in appendix C) that despite the good agreement found by BR09 between their PI+ (which diagnostically accounts for unbalanced flow) and the simulated maximum wind speed, there is a large discrepancy between the value of PI+ and that which could be obtained by substituting an estimate of the degree of unbalanced flow back into the original BR09 equation. The authors then use this supposed discrepancy as the basis for disputing that superintensity can already be largely explained by supergradient flow. However, for several reasons, the authors' analysis and argument here is fundamentally flawed. Most importantly, the authors are using an \underline{approximate} relationship between the degree of imbalance and the original PI+ equation to manipulate PI+ into a new form to make a  \underline{quantitative} comparison. But this approximate expression is only used by BR09 to make a qualitative interpretation of the influence of inertial terms on supergradient flow, and is \underline{not} used in the derivation of PI+. The correct form of PI+ as derived by BR09 agrees well with the actual simulated wind speed maximum across a wide range of simulations, and the apparent discrepancy found by the authors is an artifact of manipulating the formulation of PI+ in a manner that makes it inaccurate. Further discussion of the problems with appendix C are provided in comments \#~17-22 below.

\vspace{0.15cm}\noindent 
{\color{black}\rm
The main point that invalidates the reviewer{\textquoteright}s reasoning is that \citet{bryan09b}{\textquoteright}s Eq.~(22) is a must for their derivations to obey, not an optional approximation. This was overlooked by \citet{bryan09b} because they neglected to consider the equation of motion for  tangential velocity, from which it is obvious. That \citet{bryan09b}{\textquoteright}s derivation conflicts with the (fundamental) Eq.~(22) proves that the derivation is incorrect.

\vspace{0.15cm}\noindent 
We added three clarifying paragraphs in the beginning of appendix~C, as follows.

\vspace{0.15cm}\noindent 
"To derive Eq.~(\ref{br}), \citet[][their Eqs.~B1 and B2]{bryan09b} considered the equations of motion for radial and vertical velocities, but omitted to consider the tangential velocity equation.  From this equation for an axisymmetric inviscid flow \citep[see, e.g., Eq.~(A10b) of][]{mpi4-jas}
\begin{equation}\nonumber
u\left[f + \frac{1}{r}\frac{\partial (r v)}{\partial r}\right] + w \frac{\partial v}{\partial z}=0,
\end{equation}
it follows that, at the point of maximum tangential wind, the radial velocity is zero. This condition may be {\it approximate} in a numerical model that {\it violates E-PI{\textquoteright}s assumption of zero viscosity at the point of maximum wind}. But it must be {\it exactly} obeyed by any theoretical derivation that, as in the analysis of \citet{bryan09b}, keeps this assumption intact.

With $u=0$, the equation of motion for radial velocity, \citet{bryan09b}{\textquoteright}s Eq.~(21), becomes their Eq.~(22)
\begin{equation}\nonumber
\eta_m w_m = \frac{v_m^2}{r_m} - \alpha \frac{\partial p}{\partial r} \equiv (1 - \mathcal{B})\frac{v_m^2}{r_m}
\end{equation}
under an additional {\it minor} assumption
that $|\eta| \equiv |\partial u/\partial z - \partial w/\partial r| \gg |\partial w/\partial r|$. The identity uses our definition of $\mathcal{B}$ (\ref{gwbg}).

From the law of conservation of mass, the latter  condition should be fulfilled when $(z_m/\Delta r_m)^2 \ll 1$, where $z_m$
is the altitude of the point of maximum tangential wind (where $u = 0$) and $\Delta r_m$ is the radial half-width of the eyewall.
Under realistic conditions, this relationship should always hold.
For example, for \citet{bryan09b}{\textquoteright}s control simulation with $z_m \simeq 1$~km and $\Delta r_m \simeq 7$~km (retrieved from their Fig.~3), 
the inaccuracy of Eq.~(\ref{br2}) due to the neglect of $\partial w/\partial r$ as compared to $\eta$ should be of the order of 2\%. This appoximately agrees with
the available information about the control simulation: $\eta_m = 0.03$~s$^{-1}$ \citep[][p.~3055]{bryan09b} and
$|\partial w/\partial r| < \Delta w/\Delta r_m = 0.001$~s$^{-1}$, where we calculated $\Delta w = w_m - w = 7.5$~m~s$^{-1}$ taking into account that $w=0.5$~m~s$^{-1}$
at the eyewall outer borders and $w_m=8$~m~s$^{-1}$ at the point of maximum wind \citep[see, respectively,][their Fig.~3 and p.~3055]{bryan09b}."

\vspace{0.15cm}\noindent 
Another point that \citet{bryan09b} and the reviewer apparently neglected to consider, is as follows. How could \citet{bryan09b}{\textquoteright}s derivation have produced an estimate of $v_m$ accurate to within a couple of percent, if one of E-PI{\textquoteright}s key equations, on which this derivation is based, errs by 50\%? It could only be if one error (in \citet{bryan09b}{\textquoteright}s Eq.~8) were compensated by another (in their derivation of PI+). See also responses to Comments 17-22 below.
}

\vspace{0.15cm}
\noindent
\textbf{Minor Comments:}

\vspace{0.15cm}
\noindent
1. P4 l78 I suggest changing {\textquotedblleft}E-PI conforms to observations{\textquotedblright} to {\textquotedblleft}E-PI generally conforms to observations and numerical simulations ...{\textquotedblright}

\vspace{0.15cm}
\noindent
2. P4 l79 
Suggest changing {\textquotedblleft}has been explained{\textquotedblright} to {\textquotedblleft}can be largely explained{\textquotedblright}.

\vspace{0.15cm}
\noindent
3. P9 l156 
Move {\textquotedblleft}(A10){\textquotedblright} to just after {\textquotedblleft}with temperature{\textquotedblright}, for clarity.

\vspace{0.15cm}
\noindent
5. P11 l213 
Insert {\textquotedblleft}their{\textquotedblright} prior to {\textquotedblleft}Fig. 4c{\textquotedblright} to avoid confusion. Similarly, insert {\textquotedblleft}their{\textquotedblright} prior to {\textquotedblleft}Fig. 4{\textquotedblright} on l215.

\vspace{0.15cm}
\noindent
11. P16 l351, p19 l404 Change {\textquotedblleft}understand{\textquotedblright} to {\textquotedblleft}explain{\textquotedblright}.

\vspace{0.15cm}
\noindent
14. P21 l467 A citation should be provided here for the modification of E-PI for surface wind speed.

\vspace{0.15cm}\noindent 
{\color{black}\rm
Comments 1, 2, 3, 5, 11, 14: Revised as suggested.
}

\vspace{0.15cm}
\noindent
4. P11 l196-199

\noindent
I don't think it is true that B can be approximated as 1 in Fig. 8 of BR09. From the figure, it is clear that there is indeed non-negligible supergradient flow at the surface. 

\vspace{0.15cm}\noindent 
{\color{black}\rm
Fig.~8 of \citet{bryan09b} shows $(1- \mathcal{B})/\mathcal{B}$. Everywhere between $r \simeq 15$~km and $r \simeq 20$~km (which harbors the radius of maximum wind),  this magnitude is confined between $-0.2$ and $+0.2$, which corresponds to zero within
the accuracy adopted in that figure. Moreover, $(1- \mathcal{B})/\mathcal{B}$ apparently changes
sign approximately at the radius of maximum wind, such that somewhere close $\mathcal{B}$ is exact unity. 
In any case, $1/\mathcal{B}$ at the surface at the radius of maximum wind is certainly less than $1.2$,
while superintensity in the study of \citet{Li20c} reaches a factor of $1.5$ (see our Fig.~\ref{fig1}b).
}

\vspace{0.15cm}\noindent 
Yes, it is substantially less than at the level of peak wind speed, but this can't be used to make a qualitative inference about whether or not the surface wind speed exceeds the surface PI in that simulation. Moreover, it certainly cannot be used to make a conclusion about whether supergradient winds explain surface superintensity in a completely \underline{different simulation} in a different study (Li et al. 2020).

\vspace{0.15cm}\noindent 
{\color{black}\rm
As we noted in the previous round, one could expect {\color{blue}\it "a conclusion about whether supergradient winds explain surface superintensity"} to have been made by \citet{Li20c} themselves, and verified by their reviewers. Surprisingly, they remained completely silent on this prominent matter. To break this silence, we indicate a possible way of approaching this problem, carefully using parentheses, the subjective mood, {\it "if indeed"} and {\it "could"}.
}

\vspace{0.15cm}
\noindent
6. P14 l264-277

\noindent
I think it needs to be made clear here that {\textquotedblleft}B{\textquotedblright} as used in these calculations is not a direct evaluation of gradient wind balance, but rather using the diagnostic approximation from BR09.

\vspace{0.15cm}\noindent 
{\color{black}\rm
How $\mathcal{B}$ is obtained is, in our opinion, accurately explained in Table~\ref{tabisab}, which is referred to from this paragraph.
}

\noindent
Although the authors are clear that {\textquotedblleft}C{\textquotedblright} is being calculated from Eq. 15 and not actually estimated directly from observations, I think the authors need to be more transparent with respect to the interpretation here. They are not actually showing that large values of C compensate for strong supergradient flow (B$<1$), because C is unknown and is essentially being estimated as a residual. So the way these calculations are done, the authors would \underline{always} find that C compensates for B and vice versa, by definition, but this is unable to actually test or confirm their hypothesis.

\vspace{0.15cm}\noindent 
{\color{black}\rm
We respectively disagree. What is shown here is that for $\mathcal{B} < 1$ we have $\mathcal{C} > 1$. This illustrates our statement that air temperature increasing towards the center can compensate the effect of supergradient winds in the PI estimate. Had we been unable to find an example of such a case, this would have remained a theoretical possibility. For example, we could have found that $\mathcal{C} < 1$. That would have meant that the temperature gradient {\it enhanced} the effect of supergradient winds rather
than {\it compensated} it.
}

\vspace{0.15cm}
\noindent
7. P15 l292, 297, Eq. 17

\noindent
Why does $T_0$ appear here? I think the authors intend this to be the temperature at the top of the boundary layer, but $T_0$ is already used to mean the outflow temperature. It{\textquoteright}s possible that the authors are using the letter {\textquotedblleft}o{\textquotedblright} for outflow and the number 0 for the temperature at the top of the boundary layer, but these are hard to distinguish and so something other than $T_0$ should be used here.

\vspace{0.15cm}\noindent
{\color{black}\rm
We have added a clarifying note "see Table~\ref{tabEPI} for other notations" after Eq.~\eqref{u}. In fact, the number "0" and italic letter "$o$" are quite different, and both are used in JAS notations, see, e.g., \citet{em88}'s Eq.~(C15), where both $T_0$ and $T_o$ are present.
}

\vspace{0.15cm}
\noindent
8. P15 l305

\noindent
It isn't clear to me why the ratio of tangential to radial winds well above the surface should affect the temperature at this level or its radial gradient. Aside from the fact that the factor of 1/u is only present in this term in (17) as a mathematical convenience so it can be written in the chosen format, this equation is derived (and using approximations applicable) only for parcels at the surface. Neither the tangential nor radial wind speed (nor their ratio) at 1 km height has any direct influence on the magnitude of surface heat fluxes.

\vspace{0.15cm}\noindent 
{\color{black}\rm
We appreciate that the reviewer has apparently accepted our proposition that radial velocity (and not just $\partial p/\partial r$) controls the adiabaticity of the surface flow,  as our Eq.~(\ref{u}) describes. We respectfully disagree that $1/u$ is just a mathematical convenience. Whatever mathematical transformations one might choose to perform over Eq.~(\ref{u}), the physical message will remain there: with increasing $u$, the relative heat input gets smaller, while the horizontal temperature gradient gets closer to dry adiabatic.

\vspace{0.15cm}\noindent 
Otherwise we agree. The temperature gradient at the top of the boundary layer is more complicated than Eq.~(\ref{u}).
Since, for our conclusions, we do not need to know the total temperature difference between the eyewall and its environment at the level of maximum wind, we removed this paragraph altogether. We are concerned with the local temperature gradient at the top of the boundary layer in the eyewall. Its calculation requires additional assumptions as in our Eqs.~(\ref{comp})-(\ref{Cb}).
}

\vspace{0.15cm}
\noindent
9. P15 l307-309

\noindent
I appreciate the clarifications and revisions that the authors have made, but I still disagree with the authors' argument here, which now seems to be that differences in the radial temperature gradient near the surface compared to at the (assumed) level of maximum wind speed reflect the combined differences in radial adiabatic expansion and vertical turbulent heat fluxes. If there were an inward decrease in temperature that were due to adiabatic expansion alone, then this would be nearly the same at the two levels, because the net pressure change with radius is nearly the same (I think the authors agree with this point). 

\vspace{0.15cm}\noindent 
{\color{black}\rm
Yes, we agree.
}

\vspace{0.15cm}\noindent 
\noindent
Surface heat fluxes will counteract the expansion cooling (yes, if the flux convergence were somehow zero at the surface, this wouldn't occur, but this is a pretty unrealistic assumption to make), 

\vspace{0.15cm}\noindent 
{\color{black}\rm
A real-world {\it zero} is just a magnitude that is negligibly small compared to something else. When $u$ is large, the heat flux convergence is small compared to the change of internal energy
of the expanding gas.
}

\vspace{0.15cm}
\noindent
and turbulent mixing within the boundary layer should allow for the surface fluxes to warm the layer above the surface as well. It isn't obvious why this would cause the radial temperature gradient to be substantially \underline{less} in magnitude at 1 km than at the surface (where at both levels it is assumed that temperature is decreasing inwards). What likely would contribute substantially to vertical variation in the radial temperature gradient is condensational heating within the upper boundary layer, and evaporative cooling (from rain falling) in the lower boundary layer. This was demonstrated in an idealized modeling framework in Kepert (2016). Ultimately, there are a number of factors which collectively contribute to the radial temperature gradient and its vertical variation, and I don't think it is possible to reliably draw conclusions from the highly simplified theoretical analysis here.

\vspace{0.15cm}\noindent 
{\color{black}\rm
As we said in response to Comment 8, we agree. We removed this paragraph. We also quoted \citet{kepert16}, who observed a temperature drop comparable to Hurricane Isabel{\textquoteright}s. At the same time, we would like to note that this argument -- that there is a multitude of very complex poorly accountable factors that determnie  $\partial T/\partial r$ at the radius of maximum wind (RMW) -- works against E-PI. For E-PI to be generally valid, all those diverse processes should be somehow conspiring to produce a $\partial T/\partial r$ conforming to Eq.~(\ref{con}).
}

\vspace{0.15cm}
\noindent
10. P17 l344

\noindent
I don't follow the authors' argument for why C should be approximately the same as the level of maximum wind speed and the surface if the surface air were saturated (which I also don't think is really true). 

\vspace{0.15cm}\noindent 
{\color{black}\rm
That the surface air in the eyewall can be saturated appears to be an established fact, examples are Hurricane Isabel 2003 and Hurricane Earl 2010. But the above statement  (about similar $\mathcal{C}$) is more general, see Eq.~(\ref{iCb}) and the accompanying discussion. It also valid for an {\it unsaturated} isothermal case with $\partial \mathcal{H}/\partial r = 0$ at RMW, an assumption made by \citet{em86}.
}

\noindent
If the surface were saturated at the eyewall, then ascent in the eyewall would result in condensational heating, and a negative radial temperature gradient between the eyewall and the region outside the eyewall at the same height (with subsaturated air outside the eyewall).

\vspace{0.15cm}\noindent 
{\color{black}\rm
If the surface is saturated in the eyewall, this means that within the eyewall $\partial \mathcal{H}/\partial r = 0$. Even in this case,
with the mixing ratio increasing towards the center, there is an additional condensational heating that would produce a negative radial temperature
gradient at the level of maximum wind (if the surface is isothermal). In any case, $\mathcal{C}_b$ at the level of maximum wind
will be larger than at the surface. However, the difference is minor, about 10\% for typical conditions. This is what we estimated 
in Eqs.~(\ref{comp})-(\ref{Cb}).

\vspace{0.15cm}\noindent 
The case considered by the reviewer, with $\partial \mathcal{H}/\partial r < 0$ at the surface, was discussed following Eq.~(\ref{iCb}).
}

\vspace{0.15cm}
\noindent
12. P18 l386-392

\noindent
Although I agree that the outflow temperatures of 150 K predicted by the authors' potential intensity formulation are unrealistic and cannot be realized, I don't follow their argument that this demonstrates or is related to a violation of the assumption in E-PI that the eyewall is approximately adiabatic. The temperature increases with height above the outflow layer \underline{because} there is a stratosphere, which is also (largely) what constrains the height of the outflow layer. In the absence of turbulent mixing and radiation, the flow within the eyewall will still be adiabatic; it is not made non-adiabatic simply by the mean temperature increasing with height above. In reality, turbulent mixing within the outflow layer likely does have an influence, but as far as I can tell, this is not the argument the authors are making here. It is not true that simply by approaching the tropopause, the flow ceases to be adiabatic.

\vspace{0.15cm}\noindent 
{\color{black}\rm
The stratosphere constrains the height of the outflow due to the fact that the adiabatically ascending air becomes negatively buoyant.
Non-zero buoyancy implies a radical increase of the temperature gradient between the ascending air and its environment -- compared
to the tropospheric motions assumed in E-PI to be neutrally buoyant. This extra temperature gradient causes warming that breaks the adiabaticity of the flow especially as the vertical velocity (responsible for expansion) diminishes. We added a clarifying note to this paragraph.

\vspace{0.15cm}\noindent 
Note also that adiabaticity of the {\it eyewall} is not sufficient for E-PI to be valid. The flow must be adiabatic all way from RMW to the point where $v = 0$ (this is how the outflow in E-PI is defined). \citet{mpi4-jas} showed that for the conventional E-PI this point must be located at a radius significantly exceeding RMW.
}

\vspace{0.15cm}
\noindent
13. P19 l402

\noindent
The authors have not made clear why the warmth of the stratosphere (which in the framework of E-PI is above the region considered) somehow causes the assumption of eyewall ascent being adiabatic in E-PI theory to be violated.

\vspace{0.15cm}\noindent 
{\color{black}\rm
Please see response to Comment 12.
}

\vspace{0.15cm}
\noindent
15. P29 l629

\noindent
Need to clarify that {\textquotedblleft}u=0{\textquotedblright} is an approximation.

\vspace{0.15cm}\noindent 
{\color{black}\rm
This matter is discussed in the revised text, please see response to the Major Comment.  It is not an approximation, but an exact relationship.
}

\vspace{0.15cm}
\noindent
16. P29 l631

\noindent
I think it needs to be made clear that BR09{\textquoteright}s Eq. 22 is intended to be an approximate relationship.

\vspace{0.15cm}\noindent 
{\color{black}\rm
In the revised text, we made it clear that \citet{bryan09b}{\textquoteright}s Eq.~(22) must be obeyed by their derivation to a high precision.
Please see response to the Major Comment.
}

\vspace{0.15cm}
\noindent
17. P29 l643, Eq. C4

\noindent
I don't understand the purpose of deriving C4. 

\vspace{0.15cm}\noindent 
{\color{black}\rm
The purpose of deriving Eq.~(\ref{B}) is to check \citet{bryan09b}{\textquoteright}s derivation for self-consistency. It is a routine procedure in theoretical analysis.
}

\vspace{0.15cm}\noindent 
C1 is already the \underline{correct} expression for the PI+ (what the authors call {\textquotedblleft}$V_B${\textquotedblright}) derived by BR09, and here, the authors are substituting the simulated Vmax to solve for PI+. This doesn't seem appropriate, as PI+ is intended to be used in \underline{comparison} to the simulated Vmax. The authors are making the assumption that PI+ is equal to the simulated Vmax here, and I don't think that is valid in the context of their derivation. 

\vspace{0.15cm}\noindent 
{\color{black}\rm
\citet{bryan09b}{\textquoteright}s expression for PI+ ($v_B$ in Eq.~(\ref{br}) in our text) is a theoretical expression for maximum velocity $v_m$ that must be valid under the same assumptions as E-PI except for the gradient wind and hydrostatic balances. The same is true for Eq.~(\ref{B}). Therefore, combining these two equations with $v_B = v_m$ is justified. Had PI+ been a correct expression, this combination would not have resulted in a discrepancy. Please see also response to the Major Comment
for more details.
}

\vspace{0.15cm}\noindent
It seems that the authors are deriving C4 because it appears to result in a singularity for certain combinations of supergradient flow and dissipative heating, but this is an artifact of the derivation. The original correct expression for PI+ as shown in C1 obviously does not have such a singularity, so it shouldn't be possible to correctly manipulate the equation to obtain one.

\vspace{0.15cm}\noindent 
{\color{black}\rm
\citet{bryan09b}{\textquoteright}s expression as shown in Eq.~(\ref{br}) is incorrect. The singularity is hidden in the interrelationship between $\eta_m$, $w_m$, $v_m$ and $r_m$. It is revealed by Eq.~(\ref{B}). Please see also response to the Major Comment for more details.
}

\vspace{0.15cm}\noindent
I also note that by using the \underline{approximation} C2 the authors could have directly substituted (1-B)*$v^2$ for r*eta*w in C1 and obtained Vb$\sim 119$~m/s, which is substantially closer to the actual maximum wind speed (109 m/s) than the 144 m/s obtained from C4. Of course, what I did here isn't really valid either since I'm using the actual maximum wind speed to evaluate an expression that is meant to be compared to itself. But this illustrates how manipulating the original PI+ can lead to \underline{different} discrepant results depending on what choices are made, and that this isn't really a proper way of assessing PI+.

\vspace{0.15cm}\noindent 
{\color{black}\rm
These {\color{blue}\it \underline{different} discrepant results} is precisely how the incorrectness of \citet{bryan09b}{\textquoteright}s PI+ is manifested.
}

\vspace{0.15cm}\noindent
The more important issue here is that BR09  \underline{never} use their Eq 22 (relating eta*w to the supergradient wind speed) in the derivation of PI+. Their Eqs 23 and 24 provide the derived form of PI+, which in \underline{no way} is dependent on their Eq 22. Instead, Eq 22 is merely an approximation used to provide \underline{qualitative} aid in interpretation of the term given by the product of azimuthal vorticity and vertical velocity. Eq 22 shows that (at the location of maximum wind speed) this product is \underline{approximately} equal to the supergradient flow (divided by radius), and this demonstrates why this term leads to an increase in wind speed. But it seems clear from BR09 that this qualitative inference has no use within the actual derivation of PI+ itself. Therefore, what the authors do here in using this relationship quantitatively in order to substitute into and change the original PI+ is not valid, and so the critique of PI+ here is mistaken.

\vspace{0.15cm}\noindent 
{\color{black}\rm
That \citet{bryan09b} did not use their Eq.~(22), does not free their derivation from the obligation to obey that equation. It is like one does not invariably use the law of energy conservation in all one{\textquoteright}s derivations, but all derivations, with no exception, must obey it. As we explain in the revised text, see also response to the Major Comment, \citet{bryan09b} neglected to consider the equation of motion for tangential velocity, which shows that under their assumptions $u = 0$ at the point of maximum wind is not an optional approximation,
but an exact relationship.
}

\vspace{0.15cm}
\noindent
18. P30 l649-652

\noindent
The authors state that BR09 conclude that their PI+ works well diagnostically for predicting the maximum wind speed based on the agreement between the maximum wind speed (109 m/s) and PI+ (107 m/s) in their control simulation. But this dramatically understates the robustness of agreement between the theoretical and simulated wind maximum, because BR09 also show in their Fig. 12 that across a wide range of horizontal mixing lengths (and hence degrees of supergradient flow) in different simulations, PI+ works similarly well. The omission of this key result from the discussion makes it appear that BR09 were basing their conclusions on a single simulation, for which the authors imply may merely be a coincidence. But the authors' argument becomes much harder to sustain when considering that PI+ works well over multiple simulations. I think the authors \underline{must} make clear that the results of BR09 are not simply from looking at a single control simulation.

\vspace{0.15cm}\noindent
{\color{black}\rm
We added the following two paragraphs to discuss \citet{bryan09b}{\textquoteright}s Fig.~12.

\vspace{0.15cm}\noindent
{\it  Along with their control simulation, \citet[][their Fig.~12]{bryan09b} reported additional results intended
to illustrate a good agreement between $v_B$ (\ref{br}) and $v_m$ across a range of horizontal mixing lengths $l_h$.
However, they did not indicate whether in these additional simulations the atmosphere is inviscid at the point of maximum wind, as it approximately is in their control simulation \citep[][p.~3050]{bryan09b}. Until this is demonstrated (which will however mean that for these simulations Eq.~(\ref{br2}) also holds with a high precision), the data of  \citet{bryan09b} cannot be interpreted as supporting their Eq.~(\ref{br}), which was derived for an inviscid atmosphere. It should also be noted that while \citet[][]{bryan09b}'s Fig.~12 indicates a very close (to within a couple of percent) agreement between  $v_B$ (\ref{br}) and  $v_m$ for $47~{\rm m} \le l_h \le 1500~{\rm m}$, subsequent analysis did not confirm such a close match but revealed that for $l_h \le 300$~m there is a 20\% superintensity unexplained by $v_B$ \citep[see][p.~1137 and their Fig.~13 for $C_k/C_d = 1$ and $l_v = 200$~m]{bryan12}.

\vspace{0.15cm}\noindent
There is another reason why the claim for generality in \citet{bryan09b}{\textquoteright}s Fig.~12 warrants caution. \citet{bryan09a}{\textquoteright}s simulations used in \citet{bryan09b}{\textquoteright}s Fig.~12 use a different {\it vertical} mixing
length $l_v = 200$~m as compared to $l_v = 100$~m in \citet{bryan09b}{\textquoteright}s  control simulation. According to \citet[][see their Fig.~2 and p.~1777]{bryan09a}, varying  $l_v$ makes virtually no impact on maximum velocity over a range from $l_v = 25$~m to $l_v = 400$~m. At the same time though, this parameter is apparently instrumental in bringing E-PI{\textquoteright}s assumptions in agreement with the numerical model.  Indeed, one of E-PI{\textquoteright}s key assumptions, \citet[][]{bryan09b}{\textquoteright}s Eq.~(8) ("the fundamental closure"),  is violated by 50\% in the control simulation with $l_v = 100$~m, but with $l_v = 200$~m it becomes almost perfectly accurate
\citep[cf. Figs.~6 and 7 of][]{bryan09b}. \citet[][]{bryan09b} tentatively attributed this to the boundary layer being "better resolved" with $l_v = 200$~m. If correct, for this to be accepted as a general explanation, an analysis over a larger range of $l_v$ would seem to be required. Whether the validity of E-PI{\textquoteright}s dissipative heating formulation hinges on a model parameter that does not matter for maximum velocity, and if yes, what the implications are, remains to be investigated.
}

}

\vspace{0.15cm}\noindent
19. P31 Footnote 5

\noindent
The authors speculate that the supposed discrepancy in the theory of PI+ by BR09 is due to BR09 substituting different simulations for some of their analyses in their study, {\textquotedblleft}for an undisclosed reason{\textquotedblright}. I think this sort of speculation is improper here when made without sufficient evidence. The authors' sole piece of evidence for their claim is that BR09 {\textquotedblleft}report slightly different vm values for their control simulation vm=108 m/s on p3046 versus vm=109 m/s on p3055.{\textquotedblright} In general, it is far more likely that such a minor difference of 1 m/s would be due to either a typo or rounding than the surreptitious use of an alternate set of simulations that are undisclosed in the study. But in this case, I think the reason for the 1 m/s difference is fairly clear, when reading BR09 carefully. On p3046, BR09 state that {\textquotedblleft}We quantify the intensity of tropical cyclones in this article by vmax, which is the average of the maximum value of v from every time step during t=4-8 days.{\textquotedblright} So the 108 m/s value comes from averaging vmax over \underline{all time steps}. Then they state {\textquotedblleft}For other analyses herein, we utilize the average state computed fromhourly output during t=4-8 days.{\textquotedblright} So the 109 m/s value comes from using  \underline{hourly output}, and there is every reason to expect that this explains this very minor difference in values.

\vspace{0.15cm}\noindent
{\color{black}\rm We trust the reviewer{\textquoteright}s informed opinion that the discrepancy in the analysis of \citet{bryan09b} is not caused by different simulations described on p.~3046 versus p.~3055. We removed this footnote.
}

\vspace{0.15cm}
\noindent
20. P31 l675

\noindent
I don't think the authors claim here is correct. The derivation of BR09 does not actually imply the authors' equation C4.

\vspace{0.15cm}\noindent 
{\color{black}\rm
As we discussed in response to the Major Comment, the derivation of \citet{bryan09b} must comply
to Eq.~(\ref{B}) with a high accuracy.
}

\vspace{0.15cm}
\noindent
21. P31 l687

\noindent
I don't understand how a factor of 1/1.5 could possibly characterize {\textquotedblleft}the ratio between the left-hand and right-hand parts{\textquotedblright} of an equation. By definition, the ratio of the LHS and RHS of an equation is 1. I assume the authors must intend this statement to have a different meaning, but what this is remains unclear to me.

\vspace{0.15cm}\noindent 
{\color{black}\rm
From our own experience, we can report that to understand this is indeed a little tricky and does require an in-depth acquaintance with the logic of \citet{bryan09b}{\textquoteright}s analysis, as well as with their Fig. 6, to which our corresponding sentence (quoted by the reviewer) referred.
Normally the ratio of the LHS and RHS of an equation is indeed 1. But when an equation does not hold, it is not unity.
Some of E-PI{\textquoteright}s equations do not hold in the control simulation of \citet{bryan09b}. \citet{bryan09b} plot separately
the right-hand sides and the left-hand sides of such equations to assess the mismatch. To clarify this context,
we revised the corresponding paragraph as follows. 

\vspace{0.15cm}\noindent 
{\it We conclude by demonstrating how our alternative expression (\ref{sr1}) can be applied to the control simulation of \citet{bryan09b}. Dissipative heating in E-PI is effectively accounted for by multiplying $\tau_s/\tau_M$ in  Eq.~(\ref{tau}) by the factor of  $a \equiv T_s/T_o > 1$, to obtain $v_E^{*2} = av_E^2$ instead of $v_E^2$ in Eq.~(\ref{vE}). However, E-PI{\textquoteright}s dissipative heating equation, i.e., \citet{bryan09b}{\textquoteright}s Eq.~(8), is strongly violated in their control simulation.  The ratio between the left-hand and right-hand sides of this equation, instead of being unity, is precisely equal to $1/a = 1/1.5$ \citep[see][p.~3049 and their Fig.~6]{bryan09b}.  This means that \citet{bryan09b}{\textquoteright}s Eq.~(8) without dissipative heating -- i.e., our relationship (\ref{tau}) -- for their control simulation is exact. In consequence, E-PI{\textquoteright}s maximum velocity diagnosed from our Eq.~(\ref{dsE2}) in \citet{bryan09b}{\textquoteright}s control simulation should be $v_E^*/\sqrt{a}$. It would be $v_E^*$ if E-PI{\textquoteright}s dissipative heating equation -- \citet{bryan09b}{\textquoteright}s Eq.~(8) -- were accurate.}
}

\vspace{0.15cm}
\noindent
22. P31 l692

\noindent
I don't follow the logic here in including a factor of 1.5 in the superintensity calculation. The simulation of BR09 \underline{does} include dissipative heating, and so therefore so should any comparison to E-PI. The factor of 1.5 associated with dissipative heating should not be counted as an additional source of superintensity here.
}

\vspace{0.15cm}\noindent 
{\color{black}\rm
Yes, it should, because, in the control simulation, E-PI{\textquoteright}s dissipative heating equation errs by the dissipative heating factor $1.5$. Please see also Comment 21.
}


\begin{thebibliography}{54}
\providecommand{\natexlab}[1]{#1}
\providecommand{\url}[1]{{\tt #1}}
\providecommand{\urlprefix}{URL }
\expandafter\ifx\csname urlstyle\endcsname\relax
  \providecommand{\doi}[1]{https://doi.org/\discretionary{}{}{}#1}\else
  \providecommand{\doi}{https://doi.org/\discretionary{}{}{}\begingroup
  \urlstyle{rm}\Url}\fi

\bibitem[{Aberson et~al.(2006)Aberson, Montgomery, Bell, and Black}]{aberson06}
Aberson, S.~D., Montgomery, M.~T., Bell, M., and Black, M.: Hurricane {Isabel}
  (2003): {New} insights into the physics of intense storms. {Part II:}
  {Extreme} localized wind, Bull. Amer. Meteor. Soc., 87, 1349--1354,
  \doi{10.1175/BAMS-87-10-1349}, 2006.

\bibitem[{Barnes and Bogner(2001)}]{barnes01}
Barnes, G.~M. and Bogner, P.~B.: Comments on {\textquotedblleft}{Surface}
  observations in the hurricane environment{\textquotedblright}, Mon. Wea.
  Rev., 129, 1267--1269, \doi{10.1175/1520-0493(2001)129<1267:COSOIT>2.0.CO;2},
  2001.

\bibitem[{Bejan(2019)}]{bejan19}
Bejan, A.: Thermodynamics of heating, Proc. Roy. Soc. London, 475A, 20180\,820,
  \doi{10.1098/rspa.2018.0820}, 2019.

\bibitem[{Bell and Montgomery(2008)}]{bell08}
Bell, M.~M. and Montgomery, M.~T.: Observed structure, evolution, and potential
  intensity of category 5 {Hurricane Isabel} (2003) from 12 to 14 {September},
  Mon. Wea. Rev., 136, 2023--2046, \doi{10.1175/2007MWR1858.1}, 2008.

\bibitem[{Bister et~al.(2011)Bister, Renno, Pauluis, and Emanuel}]{bister11}
Bister, M., Renno, N., Pauluis, O., and Emanuel, K.: Comment on {Makarieva}
  {\it et al.} {\textquoteleft}{A} critique of some modern applications of the
  {Carnot} heat engine concept: the dissipative heat engine cannot
  exist{\textquoteright}, Proc. Roy. Soc. London, 467A, 1--6,
  \doi{10.1098/rspa.2010.0087}, 2011.

\bibitem[{Bryan(2012)}]{bryan12}
Bryan, G.~H.: Effects of surface exchange coefficients and turbulence length
  scales on the intensity and structure of numerically simulated hurricanes,
  Mon. Wea. Rev., 140, 1125--1143, \doi{10.1175/MWR-D-11-00231.1}, 2012.

\bibitem[{Bryan and Rotunno(2009{\natexlab{a}})}]{bryan09a}
Bryan, G.~H. and Rotunno, R.: The maximum intensity of tropical cyclones in
  axisymmetric numerical model simulations, Mon. Wea. Rev., 137, 1770--1789,
  \doi{10.1175/2008MWR2709.1}, 2009{\natexlab{a}}.

\bibitem[{Bryan and Rotunno(2009{\natexlab{b}})}]{bryan09b}
Bryan, G.~H. and Rotunno, R.: Evaluation of an analytical model for the maximum
  intensity of tropical cyclones, J. Atmos. Sci., 66, 3042--3060,
  \doi{10.1175/2009JAS3038.1}, 2009{\natexlab{b}}.

\bibitem[{Bryan and Rotunno(2009{\natexlab{c}})}]{bryan09c}
Bryan, G.~H. and Rotunno, R.: The influence of near-surface, high-entropy air
  in hurricane eyes on maximum hurricane intensity, J. Atmos. Sci., 66,
  148--158, \doi{10.1175/2008JAS2707.1}, 2009{\natexlab{c}}.

\bibitem[{Camp and Montgomery(2001)}]{CampMontgomery01}
Camp, J.~P. and Montgomery, M.~T.: Hurricane maximum intensity: {Past} and
  present, Mon. Wea. Rev., 129, 1704--1717,
  \doi{10.1175/1520-0493(2001)129<1704:HMIPAP>2.0.CO;2}, 2001.

\bibitem[{Cione et~al.(2000)Cione, Black, and Houston}]{cione00}
Cione, J.~J., Black, P.~G., and Houston, S.~H.: Surface observations in the
  hurricane environment, Mon. Wea. Rev., 128, 1550--1561,
  \doi{10.1175/1520-0493(2000)128<1550:SOITHE>2.0.CO;2}, 2000.

\bibitem[{DeMaria and Kaplan(1994)}]{demaria94}
DeMaria, M. and Kaplan, J.: Sea surface temperature and the maximum intensity
  of {Atlantic} tropical cyclones, J. Climate, 7, 1324--1334,
  \doi{10.1175/1520-0442(1994)007<1324:SSTATM>2.0.CO;2}, 1994.

\bibitem[{Emanuel(2004)}]{emanuel04}
Emanuel, K.: Tropical cyclone energetics and structure, pp. 165--192, Cambridge
  University Press, \doi{10.1017/CBO9780511735035.010}, 2004.

\bibitem[{Emanuel(2006)}]{emanuel2006}
Emanuel, K.: {Hurricanes: Tempests} in a greenhouse, Physics Today, 59, 74--75,
  \doi{10.1063/1.2349743}, 2006.

\bibitem[{Emanuel(2020)}]{emanuel20}
Emanuel, K.: The relevance of theory for contemporary research in atmospheres,
  oceans, and climate, AGU Advan., 1, \doi{10.1029/2019AV000129}, 2020.

\bibitem[{Emanuel and Rotunno(2011)}]{emanuel11}
Emanuel, K. and Rotunno, R.: Self-stratification of tropical cyclone outflow.
  {Part I: Implications} for storm structure, J. Atmos. Sci., 68, 2236--2249,
  \doi{10.1175/JAS-D-10-05024.1}, 2011.

\bibitem[{Emanuel and Rousseau-Rizzi(2020)}]{er20}
Emanuel, K. and Rousseau-Rizzi, R.: Reply to {\textquotedblleft}{Comments on}
  {\textquoteleft}{An} evaluation of hurricane superintensity in axisymmetric
  numerical models{\textquoteright}{\textquotedblright}, J. Atmos. Sci., 77,
  3977--3980, \doi{10.1175/JAS-D-20-0199.1}, 2020.

\bibitem[{Emanuel(1986)}]{em86}
Emanuel, K.~A.: An air-sea interaction theory for tropical cyclones. {Part I:
  Steady-state} maintenance, J. Atmos. Sci., 43, 585--604,
  \doi{10.1175/1520-0469(1986)043<0585:AASITF>2.0.CO;2}, 1986.

\bibitem[{Emanuel(1988)}]{em88}
Emanuel, K.~A.: The maximum intensity of hurricanes, J. Atmos. Sci., 45,
  1143--1155, \doi{10.1175/1520-0469(1988)045<1143:TMIOH>2.0.CO;2}, 1988.

\bibitem[{Emanuel(1989)}]{em89}
Emanuel, K.~A.: The finite-amplitude nature of tropical cyclogenesis, J. Atmos.
  Sci., 46, 3431--3456, \doi{10.1175/1520-0469(1989)046<3431:TFANOT>2.0.CO;2},
  1989.

\bibitem[{Emanuel(1991)}]{em91}
Emanuel, K.~A.: The Theory of Hurricanes, Annu. Rev. Fluid Mech., 23, 179--196,
  \doi{10.1146/annurev.fl.23.010191.001143}, 1991.

\bibitem[{Emanuel(1995)}]{em95}
Emanuel, K.~A.: Sensitivity of tropical cyclones to surface exchange
  coefficients and a revised steady-state model incorporating eye dynamics, J.
  Atmos. Sci., 52, 3969--3976,
  \doi{10.1175/1520-0469(1995)052<3969:SOTCTS>2.0.CO;2}, 1995.

\bibitem[{Frank(1977)}]{frank77}
Frank, W.~M.: The structure and energetics of the tropical cyclone {I. Storm}
  structure, Mon. Wea. Rev., 105, 1119--1135,
  \doi{10.1175/1520-0493(1977)105<1119:TSAEOT>2.0.CO;2}, 1977.

\bibitem[{Garner(2015)}]{garner15}
Garner, S.: The relationship between hurricane potential intensity and {CAPE},
  J. Atmos. Sci., 72, 141--163, \doi{10.1175/JAS-D-14-0008.1}, 2015.

\bibitem[{Kepert et~al.(2016)Kepert, Schwendike, and Ramsay}]{kepert16}
Kepert, J.~D., Schwendike, J., and Ramsay, H.: Why is the tropical cyclone
  boundary layer not {\textquotedblleft}well mixed{\textquotedblright}?, J.
  Atmos. Sci., 73, 957--973, \doi{10.1175/JAS-D-15-0216.1}, 2016.

\bibitem[{Kieu(2015)}]{kieu15}
Kieu, C.: Revisiting dissipative heating in tropical cyclone maximum potential
  intensity, Quart. J. Roy. Meteor. Soc., 141, 2497--2504,
  \doi{10.1002/qj.2534}, 2015.

\bibitem[{Kieu and Moon(2016)}]{kieu2016}
Kieu, C.~Q. and Moon, Z.: Hurricane intensity predictability, Bull. Amer.
  Meteor. Soc., 97, 1847--1857, \doi{10.1175/BAMS-D-15-00168.1}, 2016.

\bibitem[{Kowaleski and Evans(2016)}]{kowaleski16}
Kowaleski, A.~M. and Evans, J.~L.: A reformulation of tropical cyclone
  potential intensity theory incorporating energy production along a radial
  trajectory, Mon. Wea. Rev., 144, 3569--3578, \doi{10.1175/MWR-D-15-0383.1},
  2016.

\bibitem[{Lackmann and Yablonsky(2004)}]{la04}
Lackmann, G.~M. and Yablonsky, R.~M.: The importance of the precipitation mass
  sink in tropical cyclones and other heavily precipitating systems, J. Atmos.
  Sci., 61, 1674--1692, \doi{10.1175/1520-0469(2004)061<1674:TIOTPM>2.0.CO;2},
  2004.

\bibitem[{Li et~al.(2020)Li, Wang, Lin, and Fei}]{Li20c}
Li, Y., Wang, Y., Lin, Y., and Fei, R.: Dependence of superintensity of
  tropical cyclones on {SST} in axisymmetric numerical simulations, Mon. Wea.
  Rev., 148, 4767--4781, \doi{10.1175/MWR-D-20-0141.1}, 2020.

\bibitem[{Makarieva et~al.(2010)Makarieva, Gorshkov, Li, and Nobre}]{dhe10}
Makarieva, A.~M., Gorshkov, V.~G., Li, B.-L., and Nobre, A.~D.: A critique of
  some modern applications of the {Carnot} heat engine concept: the dissipative
  heat engine cannot exist, Proc. Roy. Soc. London, 466A, 1893--1902,
  \doi{10.1098/rspa.2009.0581}, 2010.

\bibitem[{Makarieva et~al.(2015)Makarieva, Gorshkov, and Nefiodov}]{pla15}
Makarieva, A.~M., Gorshkov, V.~G., and Nefiodov, A.~V.: Empirical evidence for
  the condensational theory of hurricanes, Phys. Lett., 379A, 2396--2398,
  \doi{10.1016/j.physleta.2015.07.042}, 2015.

\bibitem[{Makarieva et~al.(2017)Makarieva, Gorshkov, Nefiodov, Chikunov, Sheil,
  Nobre, and Li}]{ar17}
Makarieva, A.~M., Gorshkov, V.~G., Nefiodov, A.~V., Chikunov, A.~V., Sheil, D.,
  Nobre, A.~D., and Li, B.-L.: Fuel for cyclones: The water vapor budget of a
  hurricane as dependent on its movement, Atmos. Res., 193, 216--230,
  \doi{10.1016/j.atmosres.2017.04.006}, 2017.

\bibitem[{Makarieva et~al.(2019)Makarieva, Gorshkov, Nefiodov, Chikunov, Sheil,
  Nobre, Nobre, and Li}]{makarieva18b}
Makarieva, A.~M., Gorshkov, V.~G., Nefiodov, A.~V., Chikunov, A.~V., Sheil, D.,
  Nobre, A.~D., Nobre, P., and Li, B.-L.: Hurricane's maximum potential
  intensity and surface heat fluxes,
  \urlprefix\url{https://arxiv.org/abs/1810.12451}, eprint arXiv: 1810.12451v2
  [physics.ao-ph], 2019.

\bibitem[{Makarieva et~al.(2020)Makarieva, Nefiodov, Sheil, Nobre, Chikunov,
  Plunien, and Li}]{makarieva20}
Makarieva, A.~M., Nefiodov, A.~V., Sheil, D., Nobre, A.~D., Chikunov, A.~V.,
  Plunien, G., and Li, B.-L.: Comments on {\textquotedblleft}{An} evaluation of
  hurricane superintensity in axisymmetric numerical
  models{\textquotedblright}, J. Atmos. Sci., 77, 3971--3975,
  \doi{10.1175/JAS-D-20-0156.1}, 2020.

\bibitem[{Makarieva et~al.(2022)Makarieva, Gorshkov, Nefiodov, Chikunov, Sheil,
  Nobre, Nobre, Plunien, and Molina}]{mpi4-jas}
Makarieva, A.~M., Gorshkov, V.~G., Nefiodov, A.~V., Chikunov, A.~V., Sheil, D.,
  Nobre, A.~D., Nobre, P., Plunien, G., and Molina, R.~D.: Water lifting and
  outflow gain of kinetic energy in tropical cyclones, J. Atmos. Sci.,
  \doi{10.1175/JAS-D-21-0172.1}, to be published, 2022.

\bibitem[{Montgomery and Smith(2020)}]{ms20}
Montgomery, M.~T. and Smith, R.~K.: Comments on {\textquotedblleft}{An}
  evaluation of hurricane superintensity in axisymmetric numerical
  models{\textquotedblright}, J. Atmos. Sci., 77, 1887--1892,
  \doi{10.1175/JAS-D-19-0175.1}, 2020.

\bibitem[{Montgomery et~al.(2006)Montgomery, Bell, Aberson, and
  Black}]{montgomery06}
Montgomery, M.~T., Bell, M.~M., Aberson, S.~D., and Black, M.~L.: Hurricane
  {Isabel} (2003): {New} insights into the physics of intense storms. {Part I:}
  {Mean} vortex structure and maximum intensity estimates, Bull. Amer. Meteor.
  Soc., 87, 1335--1347, \doi{10.1175/BAMS-87-10-1335}, 2006.

\bibitem[{Montgomery et~al.(2019)Montgomery, Persing, and
  Smith}]{montgomery2019}
Montgomery, M.~T., Persing, J., and Smith, R.~K.: On the hypothesized outflow
  control of tropical cyclone intensification, Quart. J. Roy. Meteorol. Soc.,
  \doi{10.1002/qj.3479}, 2019.

\bibitem[{Pauluis(2011)}]{pa11}
Pauluis, O.: Water vapor and mechanical work: {A} comparison of {Carnot} and
  steam cycles, J. Atmos. Sci., 68, 91--102, \doi{10.1175/2010JAS3530.1}, 2011.

\bibitem[{Persing and Montgomery(2003)}]{persing2003}
Persing, J. and Montgomery, M.~T.: Hurricane superintensity, J. Atmos. Sci.,
  60, 2349--2371, \doi{10.1175/1520-0469(2003)060<2349:HS>2.0.CO;2}, 2003.

\bibitem[{Rotunno and Emanuel(1987)}]{ro87}
Rotunno, R. and Emanuel, K.~A.: An air-sea interaction theory for tropical
  cyclones. {Part II}: {Evolutionary} study using a nonhydrostatic axisymmetric
  numerical model, J. Atmos. Sci., 44, 542--561,
  \doi{10.1175/1520-0469(1987)044<0542:AAITFT>2.0.CO;2}, 1987.

\bibitem[{Rousseau-Rizzi and Emanuel(2019)}]{re19}
Rousseau-Rizzi, R. and Emanuel, K.: An evaluation of hurricane superintensity
  in axisymmetric numerical models, J. Atmos. Sci., 76, 1697--1708,
  \doi{10.1175/JAS-D-18-0238.1}, 2019.

\bibitem[{Rousseau-Rizzi and Emanuel(2020)}]{re20}
Rousseau-Rizzi, R. and Emanuel, K.: Reply to {\textquotedblleft}{Comments} on
  {\textquoteleft}{An} evaluation of hurricane superintensity in axisymmetric
  numerical models{\textquoteright}{\textquotedblright}, J. Atmos. Sci., 77,
  1893--1896, \doi{10.1175/JAS-D-19-0248.1}, 2020.

\bibitem[{Smith(2007)}]{smith2007}
Smith, R.~K.: Corrigendum and addendum: {A} balanced axisymmetric vortex in a
  compressible atmosphere, Tellus, 59A, 785--786,
  \doi{10.1111/j.1600-0870.2007.00258.x}, 2007.

\bibitem[{Smith and Montgomery(2013)}]{smith13}
Smith, R.~K. and Montgomery, M.~T.: How important is the isothermal expansion
  effect in elevating equivalent potential temperature in the hurricane inner
  core?, Quart. J. Roy. Meteorol. Soc., 139, 70--74, \doi{10.1002/qj.1969},
  2013.

\bibitem[{Tang and Emanuel(2010)}]{tang10}
Tang, B. and Emanuel, K.: Midlevel ventilation{\textquoteright}s constraint on
  tropical cyclone intensity, J. Atmos. Sci., 67, 1817--1830,
  \doi{10.1175/2010JAS3318.1}, 2010.

\bibitem[{Tao et~al.(2020{\natexlab{a}})Tao, Bell, Rotunno, and van
  Leeuwen}]{tao20}
Tao, D., Bell, M., Rotunno, R., and van Leeuwen, P.~J.: Why do the maximum
  intensities in modeled tropical cyclones vary under the same environmental
  conditions?, Geophys. Res. Lett., 47, \doi{10.1029/2019GL085980},
  2020{\natexlab{a}}.

\bibitem[{Tao et~al.(2020{\natexlab{b}})Tao, Rotunno, and Bell}]{tao20b}
Tao, D., Rotunno, R., and Bell, M.: Lilly{\textquoteright}s model for
  steady-state tropical cyclone intensity and structure, J. Atmos. Sci., 77,
  3701--3720, \doi{10.1175/JAS-D-20-0057.1}, 2020{\natexlab{b}}.

\bibitem[{Venkat~Ratnam et~al.(2016)Venkat~Ratnam, Ravindra~Babu, Das, Basha,
  Krishnamurthy, and Venkateswararao}]{ratnam2016}
Venkat~Ratnam, M., Ravindra~Babu, S., Das, S.~S., Basha, G., Krishnamurthy,
  B.~V., and Venkateswararao, B.: Effect of tropical cyclones on the
  stratosphere--troposphere exchange observed using satellite observations over
  the north {Indian Ocean}, Atmos. Chem. Phys., 16, 8581--8591,
  \doi{10.5194/acp-16-8581-2016}, 2016.

\bibitem[{Wang and Lin(2020)}]{wang20}
Wang, D. and Lin, Y.: Size and structure of dry and moist reversible tropical
  cyclones, J. Atmos. Sci., 77, 2091--2114, \doi{10.1175/JAS-D-19-0229.1},
  2020.

\bibitem[{Wang and Lin(2021)}]{wang21}
Wang, D. and Lin, Y.: Potential role of irreversible moist processes in
  modulating tropical cyclone surface wind structure, J. Atmos. Sci., 78, 709
  -- 725, \doi{10.1175/JAS-D-20-0192.1}, 2021.

\bibitem[{Wang et~al.(2014)Wang, Camargo, Sobel, and Polvani}]{wang2014}
Wang, S., Camargo, S.~J., Sobel, A.~H., and Polvani, L.~M.: Impact of the
  tropopause temperature on the intensity of tropical {cyclones: An} idealized
  study using a mesoscale model, J. Atmos. Sci., 71, 4333--4348,
  \doi{10.1175/JAS-D-14-0029.1}, 2014.

\bibitem[{Zhou et~al.(2017)Zhou, Held, and Garner}]{zhou2017}
Zhou, W., Held, I.~M., and Garner, S.~T.: Tropical cyclones in rotating
  radiative-convective equilibrium with coupled {SST}, J. Atmos. Sci., 74,
  879--892, \doi{10.1175/JAS-D-16-0195.1}, 2017.

\end{thebibliography}

\end{document}